\documentclass[11pt, a4paper]{article}
\usepackage{jheppub}
\usepackage{bbm}
\usepackage{multirow}
\usepackage{amsmath}
\usepackage{amssymb}
\usepackage{amsfonts}
\usepackage{mathrsfs}
\usepackage{graphicx}
\usepackage{grffile}
\usepackage[11pt]{moresize}
\usepackage[normalem]{ulem}
\usepackage[dvipsnames]{xcolor}
\usepackage[]{hyperref}
\usepackage{orcidlink}
\usepackage{relsize}
\usepackage{placeins}
\usepackage{makecell}
\usepackage{lineno}
\usepackage{mathtools}
\usepackage{halloweenmath}

\usepackage{pgfplots}
\usepackage[nomessages]{fp}

\def\ket#1{|{#1}\rangle}

\newcommand\widebar[1]{\hbox{\vbox{\hrule height 0.5pt\kern0.5ex\hbox{\kern-0.15ex\ensuremath{#1}\kern0.15em}}}}

\hypersetup %
{
 bookmarks  = true,           % show bookmarks bar?
 unicode    = false,          % non-Latin characters in Acrobat?s bookmarks
 pdfcreator = {RevTeX},       % documentclass
 colorlinks = true,           % false: boxed links; true: colored links
 linkcolor  = blue,           % color of internal links
 citecolor  = blue,           % color of links to bibliography
 filecolor  = black,          % color of file links
 urlcolor   = blue,           % color of external links
}
\newcommand{\capdef}{}
\newcommand{\mycaption}[2][\capdef]{\renewcommand{\capdef}{#2}
\caption[#1]{{\footnotesize #2}}}
\makeatletter
\renewcommand{\fnum@table}{\textbf{\tablename~\thetable}}
\renewcommand{\fnum@figure}{\textbf{\figurename~\thefigure}}
\makeatother

\newenvironment{conditions*}
{\par\vspace{\abovedisplayskip}\noindent
  \tabularx{\columnwidth}{>{$}l<{$} @{${}={}$} >{\raggedright\arraybackslash}X}}
{\endtabularx\par\vspace{\belowdisplayskip}}
\date{\today}
\preprint{IP/BBSR/2023-06}

\title{Constraining non-unitary neutrino mixing using matter effects in atmospheric neutrinos at INO-ICAL}

\author[a, b]{Sadashiv Sahoo\orcidlink{0000-0001-6719-7723},}
\author[a, b]{Sudipta Das\orcidlink{0000-0002-5508-7751},}
\author[a, b]{Anil Kumar\orcidlink{0000-0002-8367-8401}}
\author[a, b, c]{\\and Sanjib Kumar Agarwalla\orcidlink{0000-0002-9714-8866}}

\affiliation[a]{Institute of Physics, Sachivalaya Marg, Sainik School Post,\\ Bhubaneswar 751005, India}
\affiliation[b]{Homi Bhabha National Institute, \\Anushakti Nagar, Mumbai 400094, India}
\affiliation[c]{Department of Physics \& Wisconsin IceCube Particle Astrophysics Center,\\ University of Wisconsin, Madison, WI 53706, U.S.A}

\emailAdd{sadashiv.sahoo@iopb.res.in}
\emailAdd{sudipta.d@iopb.res.in}
\emailAdd{anil.k@iopb.res.in}
\emailAdd{sanjib@iopb.res.in}

\abstract{%
The mass-induced neutrino oscillation is a well established phenomenon that is based on the unitary mixing among three light active neutrinos. Remarkable precision on neutrino mixing parameters over the last decade or so has opened up the prospects for testing the possible non-unitarity of the standard 3$\nu$ mixing matrix, which may arise in the seesaw extensions of the Standard Model due to the admixture of three light active neutrinos with heavy isosinglet neutrinos. Because of this non-unitary neutrino mixing (NUNM), the oscillation probabilities among the three active neutrinos would be altered as compared to the probabilities obtained assuming a unitary 3$\nu$ mixing matrix. In such a NUNM scenario, neutrinos can experience an additional matter effect due to the neutral current interactions with the ambient neutrons. Atmospheric neutrinos having access to a wide range of energies and baselines can experience a significant modifications in Earth's matter effect due to NUNM. In this paper, we study in detail how the NUNM parameter $\alpha_{32}$ affects the muon neutrino and antineutrino survival probabilities in a different way. Then, we place a comparable and complementary constraint on $\alpha_{32}$ in a model independent fashion using the proposed 50 kt magnetized Iron Calorimeter (ICAL) detector under the India-based Neutrino Observatory (INO) project, which can efficiently detect the atmospheric $\nu_\mu$ and $\bar\nu_\mu$ separately in the multi-GeV energy range. Further, we discuss the advantage of charge identification capability of ICAL and the impact of uncertainties in oscillation parameters while constraining $\alpha_{32}$. We also compare the $\alpha_{32}$ sensitivity of ICAL with that of future long-baseline experiments DUNE and T2HK in isolation and combination.
}%

\arxivnumber{2309.16942}
\begin{document}
\maketitle
\flushbottom
%=============================%

%=============================%
\section{Introduction and motivation}
\label{sec:intro}
%=============================%
The Standard Model (SM) of particle physics is one of the most successful models that describe the interactions of the most fundamental building blocks of our observable universe~\cite{Workman:2022ynf}. It has unambiguously established the connection between the local gauge symmetries and the three fundamental force carriers (spin-1 bosons): gluons ($g$) for strong, $W^\pm$, and $Z^0$ for weak and photons ($\gamma$) for electromagnetic interactions, under the $SU(3)_C\, \times\, SU(2)_L\, \times\, U(1)_Y$ representation of the gauge group, where $C$ stands for color-charge, $L$ corresponds to the weak isospin, and $Y$ denotes the quantum number weak hypercharge. The fundamental principles of gauge invariance prescribe that all the terms in the Langragian, including the mass terms, must obey the local gauge symmetry. Since the SM does not contain right-handed neutrinos, as a consequence, neutrinos are massless in the tree-level SM Langragian. Even at the loop level, the SM content fails to generate neutrino mass due to the violation of the total lepton number by two units. However, the compelling evidence of neutrino oscillations from several pioneering experiments involving solar~\cite{Super-Kamiokande:1998oic, Super-Kamiokande:2001bfk, SNO:2001kpb, Super-Kamiokande:2002ujc, Super-Kamiokande:2005wtt, Super-Kamiokande:2008ecj, Super-Kamiokande:2010tar, SNO:2011hxd}, reactor~\cite{Eguchi:2002dm, Araki:2004mb, An:2012eh, Ahn:2012nd, KamLAND:2013rgu, RENO:2018dro, DayaBay:2018yms, DoubleChooz:2019qbj}, atmospheric~\cite{Achar:1965ova, Fukuda:1998mi, Ashie2004, IceCube:2014flw, Super-Kamiokande:2017yvm}, and accelerator~\cite{K2K:2004iot, Adamson:2008zt, MINOS:2013xrl, MINOS:2013utc, T2K:2019bcf, NOvA:2019cyt} neutrinos indicate that they must have non-zero non-degenerate tiny masses and should mix with each other. This novel phenomenon of neutrino-flavor transition can be effectively parameterized in terms of the three mixing angles ($\theta_{12}$, $\theta_{13}$ and $\theta_{23}$), two independent mass-squared splittings ($\Delta m^2_{21}$ and $\Delta m^2_{31}$), and one leptonic Dirac CP phase ($\delta_{\rm CP}$). After a remarkable discovery of neutrino oscillation phenomena~\cite{Kajita:2016cak, McDonald:2016ixn}, we are now in the precision era of neutrino physics where only a few oscillation parameters are yet to be measured precisely, such as $\delta_{\rm CP}$, the octant of $\theta_{23}$, and the neutrino mass ordering, i.e., the sign of $\Delta m^2_{31}$. Neutrinos are massless in the SM, and the exclusive lab-based evidence of the non-zero neutrino mass as required by neutrino oscillation indicates the pressing need for theories beyond the Standard Model (BSM) to accommodate non-zero neutrino mass and mixing. Therefore, neutrino oscillation can act as a unique probe to study various BSM scenarios.

The precision data on the Z-decay width at the $e^+e^-$ collider at LEP suggest that there can be only three light active flavor neutrinos i.e., $N_{\nu_{\rm active}} = 2.9840\,\pm\,0.0082$~\cite{ALEPH:2005ab}. However, there are some anomalous results from the experiments such as LSND~\cite{LSND:1995lje, LSND:2001aii}, MiniBooNE~\cite{MiniBooNE:2007uho, MiniBooNE:2010idf},  Gallium radioactive source experiments GALLEX~\cite{Kaether:2010ag}, SAGE~\cite{Abdurashitov:2009tn}, and BEST~\cite{Barinov:2022wfh}, and Neutrino-4~\cite{NEUTRINO-4:2018huq}  point towards oscillations with a significantly large mass-squared difference of $\Delta m^2 \sim1 \text{ eV}^2$ as compared to the standard solar and atmospheric mass-squared splittings. These anomalous results observed at the short-baseline experiments suggest the existence of a fourth neutrino mass eigenstate $\nu_4$ at the eV-scale which has to be gauge singlet because of the bounds on the number of weakly-interacting light neutrino state from the LEP experiment~\cite{ALEPH:2005ab}. This gauge-singlet neutrino known as the ``sterile'' neutrino can reveal its existence via active-sterile mixing. Also, an unidentified emission line in the X-ray spectra of galaxy clusters is observed by the XMM-Newton and Chandra space telescopes~\cite{Boyarsky:2014jta, Bulbul:2014sua}. It is proposed that this X-ray emission line may originate due to the decay of a resonantly-produced keV-scale sterile neutrino dark matter~\cite{Abada:2014zra,Abazajian:2014gza,Ng:2015gfa,Schneider:2016uqi}. These heavy sterile neutrinos appear naturally in various BSM scenarios which are responsible for generating small neutrino masses\footnote{Several BSM possibilities can be introduced in the framework of effective field theory such as the effective lepton-number-violating dimension-five Weinberg operator which may account for the small neutrino masses~\cite{Weinberg:1979sa,Weinberg:1980bf}.} (e.g. seesaw)~\cite{Minkowski:1977sc, Yanagida:1979as, Gell-Mann:1979vob, Ramond:1979py, Yanagida:1980xy, Schechter:1980gr, Mohapatra:1980yp, Schechter:1981cv,Ma:1998dn} and their admixture with the three light active neutrinos may cause deviations from unitarity of the standard 3$\nu$ mixing matrix. This unitarity violation of the lepton mixing matrix can affect the oscillation probabilities of three active neutrinos~\cite{Antusch:2006vwa,Antusch:2009pm,Escrihuela:2015wra,Fong:2017gke}.

In this paper, we study in detail the impact of NUNM in three-neutrino oscillations in a model independent fashion in the context of the proposed 50 kt iron calorimeter (ICAL) detector at the India-based Neutrino Observatory (INO)~\cite{ICAL:2015stm} using atmospheric neutrinos. Here, we show how the atmospheric neutrinos having access to a wide range of baselines and energies, can feel the presence of NUNM during their oscillations. The ICAL detector is designed to have a magnetic field of 1.5 T~\cite{Behera:2014zca} which enables ICAL to distinguish between $\mu^-$ and $\mu^+$ events produced from the interactions of atmospheric $\nu_\mu$ and $\bar{\nu}_\mu$, respectively. This charge identification (CID) capability helps ICAL to achieve its primary goal of determining the neutrino mass ordering\footnote{Normal mass ordering (NMO) scenario corresponds to $m_1\, <\, m_2\, < m_3$ and inverted mass ordering (IMO) denotes the situation where $m_3\, <\, m_1\, < m_2$.} by observing atmospheric $\nu_\mu$ and $\bar\nu_\mu$ separately in the multi-GeV energy range over a wide range of baselines~\cite{Devi:2014yaa}. The ICAL has excellent detector resolutions to measure momenta and directionality of $\mu^-$ and $\mu^+$~\cite{Chatterjee:2014vta}. It can also measure the energies of hadron showers that are produced during atmospheric neutrino interactions~\cite{Devi:2013wxa}. Using the ICAL detector response, the ICAL collaboration has performed a plethora of studies to measure the standard three-flavor oscillation parameters as well as to probe several BSM scenarios~\cite{GOSWAMI2009198, Ghosh:2012px, Thakore:2013xqa, Ghosh:2013mga, Devi:2014yaa, Dash:2014fba, Chatterjee:2014oda, Choubey:2015xha, Mohan:2016gxm, Behera:2016kwr, Khatun:2017adx, Choubey:2017vpr, Choubey:2017eyg, Kaur:2017dpd, Rebin:2018fdl, Thakore:2018lgn, Khatun:2018lzs, Tiwari:2018gxz, Datta:2019uwv, Dar:2019mnk, Khatun:2019tad, Kumar:2020wgz, Kumar:2021faw, Kumar:2021lrn, Sahoo:2021dit, Sahoo:2021pkr, Upadhyay:2021kzf, Sahoo:2022rns, Senthil:2022tmj, Upadhyay:2022jfd, Raikwal:2022nqk, Raikwal:2023lzk}. Using $500$ kt$\cdot$yr exposure of ICAL, we study its performance in unraveling the signatures of NUNM and demonstrate the importance of its CID capability. 

We start this paper with a qualitative description of the NUNM scenario and its implications on the neutrino propagation Hamiltonian in section~\ref{sec:Form}. In section~\ref{sec:atm-tool}, we discuss the impact of NUNM  on the survival probabilities of atmospheric $\nu_\mu$ and $\bar\nu_\mu$. In section~\ref{sec:ICAL-detector}, we explain the detector configuration, simulation methodology, and events reconstruction. Section~\ref{sec:Stat_Meth} deals with the statistical methodology that we use while calculating the sensitivity of ICAL towards NUNM in terms of $\Delta\chi^2$. Then, in section~\ref{sec:Results}, we show the sensitivities obtained using the $500$ kt$\cdot$yr exposure of ICAL and discuss the impact of uncertainties of oscillation parameters, the true value of atmospheric mixing angle $\theta_{23}$, and the advantage of using CID capability of ICAL. Further, we perform a quantitative comparison between the sensitivities towards NUNM obtained using ICAL and the next-generation long-baseline (LBL) experiments: Deep Underground Neutrino Experiment (DUNE)~\cite{DUNE:2015lol,DUNE:2020lwj,DUNE:2020ypp,DUNE:2020jqi,DUNE:2020fgq,DUNE:2021mtg,DUNE:2021cuw} and Tokai to Hyper-Kamiokande (T2HK)~\cite{Abe:2016srs,Abe:2018ofw}. In appendix~\ref{app:A1}, we provide a brief discussion on the lower-triangular formulation of the NUNM matrix in the oscillation framework. Finally in appendix~\ref{app:A2}, we derive approximate analytical expression of $\nu_\mu$ survival probability in the presence of NUNM which helps us to explain the role of neutral current (NC) matter effect while probing the NUNM scenario using atmospheric neutrinos. 

%=============================%
\section{Formalism of non-unitary neutrino mixing (NUNM)}
\label{sec:Form}
%=============================%
In the phenomena of mass-induced neutrino flavor oscillations, the weak flavor eigenstates, $\ket{\nu_\beta}$ where $\beta \in \{e,\,\mu,\,\tau\}$, are a linear superposition of the mass eigenstates, $\ket{\nu_j}$ where $j \in \{1,\,2,\,3\}$. The flavor and mass eigenstates are related to each other with the help of a $3\times3$ unitary mixing matrix $U$ in the following fashion:
\begin{align}
\ket{\nu_\beta} & = \sum_{j\,=\,1}^{3} U^\ast_{\beta j}\, \ket{\nu_j}\,.
\label{Eq:2.1}
\end{align}
In the above equation, the mixing matrix $U$ is known in literature as the Pontecorvo-Maki-Nakagawa-Sakata (PMNS) matrix~\cite{Pontecorvo:1957qd, Pontecorvo:1967fh, Maki:1962mu}. Considering the plane wave approximation of the ultra-relativistic neutrino states,\footnote{For a detailed quantum mechanical description on mass-induced neutrino flavor transitions, see refs.~\cite{Giunti:2002xg, Giunti:2007ry, Akhmedov:2009rb, Akhmedov:2019iyt}.} the oscillation probability ($P_{\eta\beta}$) can be expressed as:
\begin{align}
P_{\eta\beta} = \delta_{\eta\beta} &- 4\,\sum_{i\, >\, j}^{3}Re\bigg[U^\ast_{\eta i}U_{\beta i}U_{\eta j}U^\ast_{\beta j}\bigg] \sin^2\left(\Delta m^2_{ij}\cdot\frac{L}{4E}\right)\nonumber\\
&+2\,\sum_{i\, >\, j}^{3}Im\bigg[U^\ast_{\eta i}U_{\beta i}U_{\eta j}U^\ast_{\beta j}\bigg]\sin\left(\Delta m^2_{ij}\cdot\frac{L}{2E}\right)\,.
\label{Eq:2.2}
\end{align}
Here, $\Delta m^2_{ij} = m^2_i - m^2_j$, and $L$ and $E$ are the propagation length and energy of neutrino, respectively. The $\eta$, $\beta$ are neutrino flavor indices, $i$, $j$ are neutrino mass indices, and $\delta_{\eta\beta}\, =\, \sum_{j} U_{\eta j}U^\ast_{\beta j}$. In eq.~(\ref{Eq:2.2}), $U\,\to\,U^\ast$ for antineutrinos. The second real term is CP-conserving, whereas the third imaginary term is CP-violating~\cite{Sato:2000wv, Nunokawa:2007qh}. Under the unitary neutrino mixing scenario, the total oscillation probability is conserved, i.e., $\sum_{\beta}\, P_{\eta\beta}\, = \, 1$. However, if NUNM is realized in Nature, the immediate consequence would be the non-conservation of the three-neutrino oscillation probabilities (see appendix \ref{app:A1} for a detailed discussion), i.e., $\sum_{\beta}\, P_{\eta\beta}\, \ne \, 1$. The scenario of NUNM would be obvious if the PMNS acts as a sub-matrix of a global unitary lepton mixing matrix ($\widetilde U_{n \times n}$) that contains the admixture of a large number of neutrino species ($n$). Such an NUNM scenario can be naturally accounted from either the consequence of the seesaw mechanism~\cite{Schechter:1981cv, Hettmansperger:2011bt} or extended light neutrino mixings~\cite{Antusch:2006vwa, Fong:2023fpt}. Assuming the group of active three light neutrinos ($\nu_1,\, \nu_2,\, \nu_3$) is the lightest among all possible neutrinos, and their mass-ordering uncertainties do not alter the status of the extra neutrino states. Then, the generalized linear mixing of neutrinos can be represented as:
%=============================%
\begin{align}
  \begin{bmatrix}
    \nu_e    \\
    \nu_\mu  \\
    \nu_\tau \\
    \vdots
  \end{bmatrix}_{n \times 1}  =
  \left(\left[\begin{array}{cccc}
    \widetilde{U}_{e1}    & \widetilde{U}_{e2}    & \widetilde{U}_{e3}    & \ldots\\
    \widetilde{U}_{\mu1}  & \widetilde{U}_{\mu2}  & \widetilde{U}_{\mu3}  & \ldots\\
    \widetilde{U}_{\tau1} & \widetilde{U}_{\tau2} & \widetilde{U}_{\tau3} & \ldots\\
    \vdots    & \vdots    & \vdots    & \ddots
  \end{array}\right]_{n \times n}
  \xRightarrow{\rm B.R.} \,
  \left[\begin{array}{cc}
  \mathcal{N}_{3\times 3}     & \mathcal{S}_{3\times (n-3)}     \\
                              &                                 \\
  \mathcal{V}_{(n-3)\times 3} & \mathcal{T}_{(n-3) \times (n-3)}
  \end{array}\right]\right)\cdot
  \begin{bmatrix}
    \nu_1 \\
    \nu_2 \\
    \nu_3 \\
    \vdots
  \end{bmatrix}_{n \times 1}\,.
  \label{Eq:2.3}
\end{align}
%=============================%
An effective block representation (B.R.) of $\widetilde U_{n\times n}$ reveals the mode of mixings among the neutrino species. Here, $\mathcal{N}$ represents the mixing among light neutrino states, while $\mathcal{S}$ and $\mathcal{V}$ are the admixture of light and heavy neutrino states, and $\mathcal{T}$ stands for the mixing of only heavy neutrinos. For the case of three light active SM neutrinos, the mixing is given by $3\times3$ matrix  $\mathcal{N}$ which is the lowest order NUNM accessible in the low-energy experiments. Therefore, it make sense to replace the $3\times3$ unitary PMNS matrix by the new  non-unitary matrix $\mathcal{N}$. In the literature, there are various ways to parameterize this NUNM matrix~\cite{Antusch:2006vwa, Xing:2012kh, Bielas:2017lok, Flieger:2019eor, Ellis:2020ehi, Hu:2020oba}. However, Okubo's notation~\cite{Okubo:1962zzc} to decompose $\mathcal{N}$ into a lower-triangular matrix ($\alpha$) and PMNS matrix ($U$) is the most convenient one for the neutrino oscillation studies (see eq.~(\ref{Eq:A1.11}) in appendix \ref{app:A1} for a detailed explanation). In this notation, $\mathcal{N}$ can be expressed as follows~\cite{Hettmansperger:2011bt, Escrihuela:2015wra, Agarwalla:2021owd, Gariazzo:2022evs, Arguelles:2022tki}:
%=============================%
\begin{align}
  \mathcal{N} & \,=\, \left(I\,+\,\alpha\right)\,\cdot\,U\,,
  \label{Eq:2.4}
\end{align}
where 
\begin{align}  
  \alpha & \,=\,
  \left(\begin{array}{ccc}
  \alpha_{11} & 0           & 0           \\
  \alpha_{21} & \alpha_{22} & 0           \\
  \alpha_{31} & \alpha_{32} & \alpha_{33}
  \end{array}\right)\,.
  \label{Eq:2.5}
\end{align}
%=============================%
In eq~(\ref{Eq:2.4}), $I$ is the identity matrix, $\alpha$ is the lower-triangular matrix and $U$ is the standard $3\times3$ PMNS matrix. In eq~(\ref{Eq:2.5}), the diagonal elements $\alpha_{ii}$ are real and close to zero. The off-diagonal elements  $\alpha_{ij}$ in eq~(\ref{Eq:2.5}) are complex in general. A detailed discussion on the properties of $\alpha$ is given in the appendix of ref.~\cite{Escrihuela:2015wra}. In this scenario, the effective Hamiltonian of ultra-relativistic left-handed neutrinos passing through the ambient matter can be written in the three-neutrino mass basis as:
%=============================%
\begin{align}
  \mathcal{H} & = \frac{1}{2E}
  \left(\begin{array}{ccc}
  0 & 0               & 0              \\
  0 & \Delta m^2_{21} & 0              \\
  0 & 0               & \Delta m^2_{31} 
  \end{array}\right)\;
  +\;\; \mathcal{N}^\dagger\cdot \sqrt{2}\,G_F
  \left(\begin{array}{cccc}
  N_e - N_n/2 & 0       & 0 \\
  0           & - N_n/2 & 0 \\
  0           & 0       & - N_n/2
  \end{array}\right)\cdot\mathcal{N}\,,
  \label{Eq:2.6}
\end{align}
%=============================%
where the first term represents the kinematics of the Hamiltonian in vacuum, and the last term includes the effective matter potentials for neutrinos induced due to both charged-current (CC) interactions, i.e., with ambient electrons, and neutral-current interactions, i.e., with ambient neutrons. The charged-current  matter potential ($V_{\rm CC}$) is quantified as $\sqrt{2}G_FN_e$, where $G_F$ is Fermi constant and $N_e$ is the electron number density of the matter. Unlike standard three-neutrino oscillations, the matter potential ($V_{\rm NC}$) due to neutral-current interactions survives in the Hamiltonian in the case of non-unitary mixing. The strength of $V_{\rm NC}$ quantifies as $-\,G_FN_n/\sqrt{2}$, where $N_n$ is the ambient neutron number density. For a given matter density $\rho$ along the neutrino trajectory inside Earth, both the CC and NC effective matter potentials can be written as:
\begin{align}
  V_{\rm CC} & \; \approx \;\;\; 7.6 \times 10^{-14}\cdot Y_e\cdot\rho\left({\rm g/cm^3}\right) \mbox{eV},\\
  V_{\rm NC} & \; \approx -\,3.8 \times 10^{-14}\cdot Y_n\cdot\rho\left({\rm g/cm^3}\right) \mbox{eV},
  \label{Eq:2.7-8}
\end{align}
where $Y_e$ and $Y_n$ are the electron and neutron number fractions in the medium, respectively.\footnote{$Y_e\,=\, N_e/(N_p\,+\,N_n)$ and $Y_n\,=\, N_n/(N_p\,+\,N_n)$ where $N_p$ is proton number density in the matter.} For antineutrinos, $\mathcal{N} \to \mathcal{N}^*$, and both the matter potentials change sign. In the NUNM scenario, there is a critical imprint of neutral-current matter effect which is proportional to the density of the medium. Therefore, the atmospheric neutrinos propagating through a high-density medium inside Earth's core can experience significant modifications in the oscillation probabilities, which in turn, can be used to probe the NUNM scenario. In this paper, we show that at  INO-ICAL, the efficient observation of atmospheric neutrinos traveling large distances inside Earth plays a crucial role in probing the possible presence of the NUNM parameter $\alpha_{32}$ which alters the $\nu_\mu$ survival probabilities significantly. 

Improved precision on the neutrino oscillation parameters over the past decade or so has opened up the possibility to test the unitarity of the PMNS matrix in the currently running and upcoming neutrino experiments. Along that direction, several studies have been performed to place bounds on the NUNM parameters using the available experimental outcomes~\cite{Blennow:2016jkn,Miranda:2019ynh,Forero:2021azc,Blennow:2023mqx}. 
For an example, using the limits on appearance probabilities placed by the short-baseline~(SBL) experiment NOMAD at CERN~\cite{NOMAD:2001xxt, NOMAD:2003mqg}, the authors in refs.~\cite{Escrihuela:2015wra, Ge:2016xya, Blennow:2016jkn} estimated the bound on the NUNM parameter $|\alpha_{32}| < 0.012$ at 95\% confidence level. In ref.~\cite{Forero:2021azc}, the authors performed a global analysis of the results obtained from the SBL experiments: NOMAD and NuTeV, and LBL experiments: MINOS/MINOS+, T2K and NO\(\nu\)A to place the limits on various NUNM parameters. They estimated a bound of $|\alpha_{32}| < 0.017$ at 99\% confidence level.\footnote{Note that while analysing the neutral-current data of MINOS/MINOS+, the authors in ref.~\cite{Forero:2021azc} used the triangular inequalities  $\alpha_{ij}\leq\sqrt{1-(1+\alpha_{ii})^2}\sqrt{1-(1+\alpha_{jj})^2}$ to derive the limit on $|\alpha_{32}|$.} Using the present constraints on active-sterile mixing~\cite{MINOS:2016viw, Super-Kamiokande:2014ndf}, the authors in ref.~\cite{Blennow:2016jkn} obtained a limit of $|\alpha_{32}|<0.053$ at 95\% C.L., which is valid for $\Delta m^2_{41}\sim 0.1-1$ eV$^2$. There, the authors also estimate a future constraint on $\alpha_{32}$ of around $\le0.3$ at 90\% C.L. using the LBL setup of DUNE~\cite{DUNE:2015lol,DUNE:2020lwj,DUNE:2020ypp,DUNE:2020jqi,DUNE:2020fgq,DUNE:2021mtg,DUNE:2021cuw}. Exploiting the synergies among the next-generation LBL experiments DUNE,  T2HK~\cite{Abe:2015zbg, Abe:2018ofw}, and T2HK with another detector in Korea~\cite{Abe:2016srs}, the authors in ref.~\cite{Agarwalla:2021owd} derive a future bound of $|\alpha_{32}| < 0.27$ at 90\% confidence level. Using the projected data from near and far detectors of the proposed ESS\ensuremath{\nu}SB long-baseline setup, the authors in ref.~\cite{Capozzi:2023ltl} estimated the improved limits on various NUNM parameters. In ref.~\cite{Celestino-Ramirez:2023zox}, the authors estimated a future bound of $|\alpha_{32}| < 0.048$ and $|\alpha_{32}| < 0.043$ at 90\% confidence level, using FASER$\nu$ and FASER$\nu$2 detector setups, respectively. Apart from neutrino oscillation experiments, the NUNM parameters are also tightly constrained by several non-oscillation experiments. In ref.~\cite{Blennow:2023mqx}, the authors placed a bound of $|\alpha_{32}|<0.011$ at 95\% C.L. in the generic unitarity violation scenario, using the current flavor and electroweak precision data. This bound becomes even more stringent while considering some specific neutrino mass models.
For an example, if the SM contents are extended with three heavy right-handed neutrinos in the low-scale seesaw models, then the electroweak data place a constraint on $|\alpha_{32}|<1.2\times 10^{-3}$ at 95\% confidence level~\cite{Blennow:2016jkn}. In our conclusion section, we show a comparison between all the existing and future limits on $|\alpha_{32}|$, and the expected bounds that we obtain in the present work using the proposed ICAL detector.\footnote{In the present study, we do not consider the triangular inequalities while estimating the future constraints on $\alpha_{32}$. We perform our analysis in a model independent fashion with one-parameter-at-a-time using the standalone INO-ICAL atmospheric neutrino setup, assuming a generic unitarity violation. Since other NUNM parameters are considered to be zero, triangular inequalities are not relevant in our analysis.}

Note that the tau neutrino row of the lepton mixing matrix is not well constrained under the assumption of NUNM. Lately, several neutrino oscillation experiments have revealed the appearance of tau neutrinos in their data sets, for an example, LBL experiment OPERA~\cite{OPERA:2018nar}, atmospheric neutrino experiments Super-Kamiokande~\cite{Super-Kamiokande:2017edb}, IceCube-DeepCore~\cite{IceCube:2019dqi}, and KM3NeT/ORCA6~\cite{Geiselbrecht:2023zfv}, and astrophysical tau neutrinos at IceCube~\cite{IceCube:2020fpi, IceCube:2023fgt}. Undoubtedly, these new data sets involving tau neutrinos have improved our knowledge of the tau neutrino part of the lepton mixing matrix~\cite{Denton:2021mso, Denton:2021rsa}, which in turn can shed light on the possible non-unitarity of the neutrino mixing matrix. Future neutrino experiments such as IceCube-Upgrade~\cite{IceCube:2023ins}, IceCube-Gen2~\cite{IceCube-Gen2:2020qha}, KM3NeT/ORCA~\cite{Eberl:2017plv}, T2HK~\cite{Abe:2018ofw}, DUNE~\cite{DeGouvea:2019kea}, and INO-ICAL~\cite{Senthil:2022tmj} are also expected to provide crucial insights on the tau neutrino matrix elements.

%=============================%
\begin{table}[t]
  \centering
  \begin{tabular}{|c|c|c|c|c|c|c|}
    \hline
    $\sin^2 2\theta_{12}$ & $\sin^2\theta_{23}$ & $\sin^2 2\theta_{13}$ & $\delta_{\rm CP}$ & $\Delta m^2_{21}$ (eV$^2$) & $\Delta m^2_
    \text{32}$ (eV$^2$) & Mass Ordering\\
    \hline
    0.855 & 0.5 & 0.0875 & 0 & $7.4\times10^{-5}$ & $2.46\times 10^{-3}$ & Normal (NMO)\\
    \hline 
  \end{tabular}
  \mycaption{The benchmark values of neutrino oscillation parameters used in this analysis, which agree with the current global fit values ~\cite{Esteban:2020cvm, NuFIT, Capozzi:2021fjo, deSalas:2020pgw}.}
  \label{Tab:1}
\end{table}
%=============================%

%=============================%
\section {Atmospheric neutrinos: a unique tool to probe NUNM}
\label{sec:atm-tool}
%=============================%

The primary cosmic rays once enter into the atmosphere of Earth, interact with air nuclei in the high-altitude atmosphere producing mainly pions and less abundantly kaons~\cite{Gaisser:2002jj}. These mesons for an example $\pi^+$ decays to a muon ($\mu^+$) and a muon neutrino ($\nu_\mu$). This secondary muon is unstable and can further decay to a positron ($e^+$), a muon antineutrino ($\bar{\nu}_\mu$), and a electron neutrino ($\nu_e$). Similarly, $\pi^-$ also goes through the charge conjugate decay chain. Tiny contribution also comes from the kaon decay. The energy of atmospheric neutrino ranges from a few MeV to more than hundreds of TeV. They arrive at the detector from all possible directions traveling a wide range of path lengths starting from an atmospheric height of 15 km (in the downward direction) to as large as the diameter of Earth (in the upward direction). It allows us to explore the impact of various new physics scenarios including the NUNM hypothesis (main thrust of this paper) on neutrino flavor oscillations for several values of $L ({\rm km})/E ({\rm GeV})$ with an emphasis in the multi-GeV energy range for a detector like ICAL.

The ICAL detector is designed to efficiently detect $\mu^-$ and $\mu^+$ events produced during the charged-current interactions of $\nu_\mu$ and $\bar{\nu}_\mu$, respectively (see discussion in section~\ref{sec:ICAL-detector}). Around 98\% of these interactions are contributed by $\nu_\mu$ and $\bar{\nu}_\mu$ disappearance channels. In the expression of survival probability of muon neutrino, $P(\nu_\mu \to \nu_\mu)$, the real component of the NUNM parameter $\alpha_{32}$ has the prominent effect at the leading order and it couples directly with the matter effect term driven by neutral-current interactions inside Earth~\cite{Agarwalla:2021owd}. Therefore, in this paper, we primarily focus on the NUNM parameter $\alpha_{32}$ and consider it to be real having both negative and positive values. Later, we also address in detail the impact of the complex phase $\phi_{32}$ associated with the NUNM parameter $\alpha_{32}$~($\equiv |\alpha_{32}|e^{-i\phi_{32}}$). Note that while focusing on $\alpha_{32}$,  all other NUNM parameters are considered to be zero. During our analysis, we observe that the above-mentioned neutral-current matter potential drives the main sensitivity towards $\alpha_{32}$ which is also evident from eq.~(\ref{Eq:3.1}).

Here, we consider three-flavor neutrino oscillations in the presence of matter with the Preliminary Reference Earth Model (PREM)~\cite{Dziewonski:1981xy} as the density profile of Earth. For simplicity, we assume Earth's matter to be neutral ($N_e = N_p$) and iso-scalar ($N_p = N_n$) where $Y_e\; \equiv \;Y_n\;\simeq\; 0.5$. We use benchmark values of six oscillation parameters as given in table \ref{Tab:1}. Note that the value of $\Delta m^2_{32}$ is calculated from the effective mass-squared difference ($\Delta m^2_{\rm eff}$)\footnote{The effective mass-squared difference $\Delta m^2_{\rm eff}$ is defined in terms of $\Delta m^2_{32}$ in the following way~\cite{deGouvea:2005hk,Nunokawa:2005nx}:
    \begin{align}
      \Delta m^2_{\rm eff} \,=\, & \Delta m^2_{32}\, + \,\Delta m^2_{21} (\sin^2\theta_{12}\, +\, \cos \delta_\text{CP} \sin\theta_{13}\sin2\theta_{12}\tan\theta_{23}).
      \label{eq:eff_dmsq-1}
    \end{align}
} whose value is taken to be $2.49 \times 10^{-3}$ eV$^2$. For NMO, we use the positive value of $\Delta m^2_{\rm eff}$, whereas for IMO, $\Delta m^2_{\rm eff}$ is taken to be negative with the same magnitude. 
%=============================%
\begin{figure}[t]
  \centering
  \includegraphics[width=0.425\textwidth]{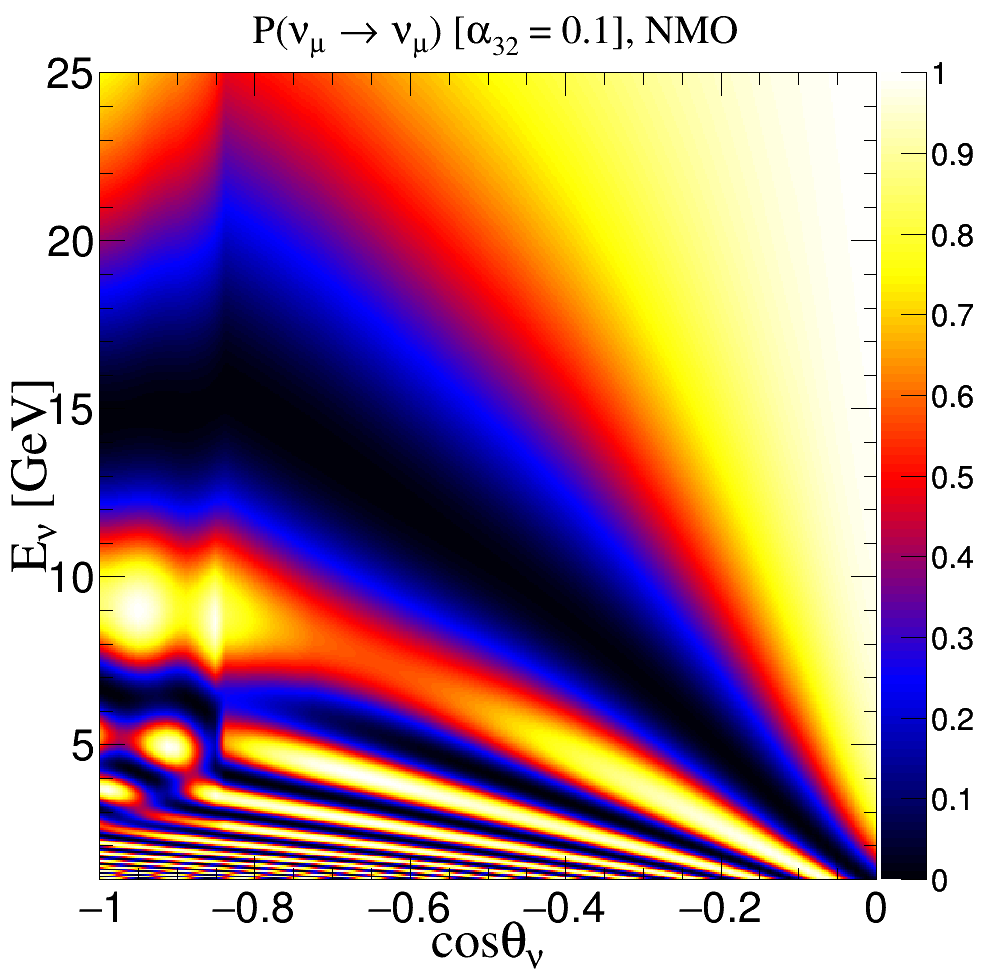}
  \hspace{0.1 cm}
  \includegraphics[width=0.425\textwidth]{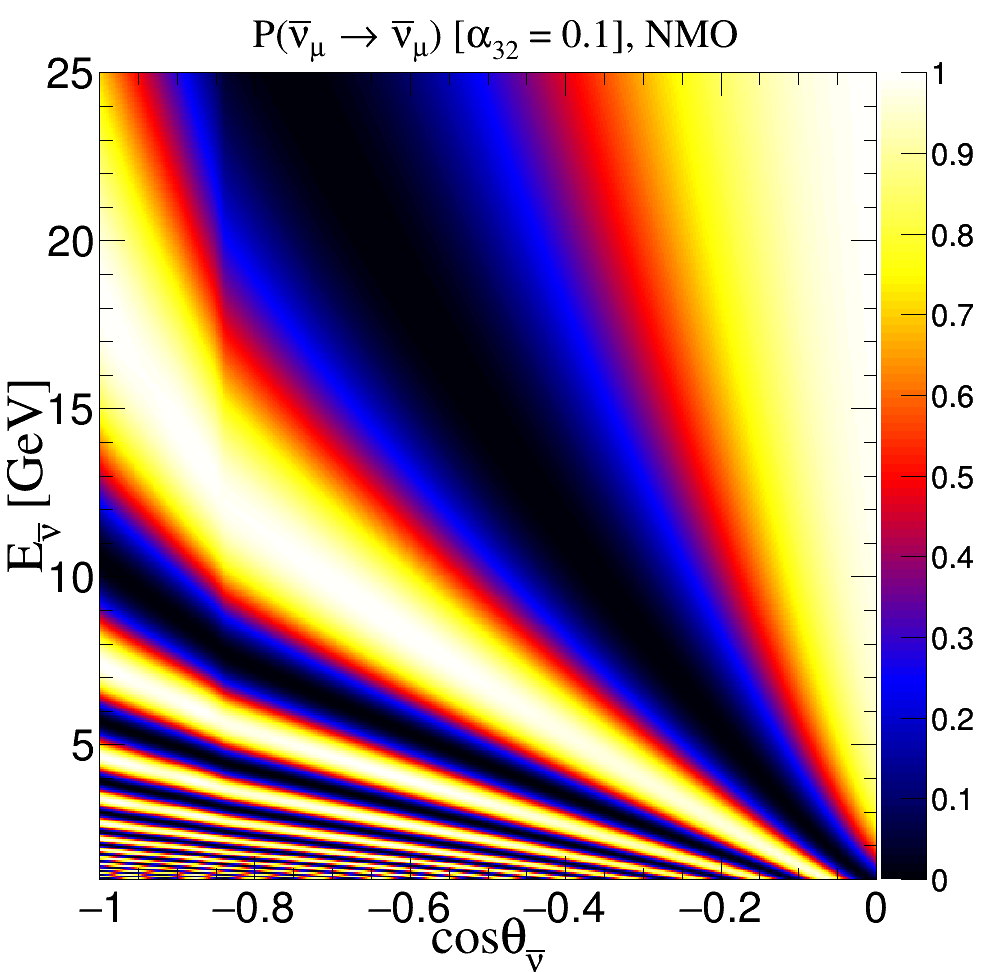}\\
  \vspace{0.1 cm}
  \includegraphics[width=0.425\textwidth]{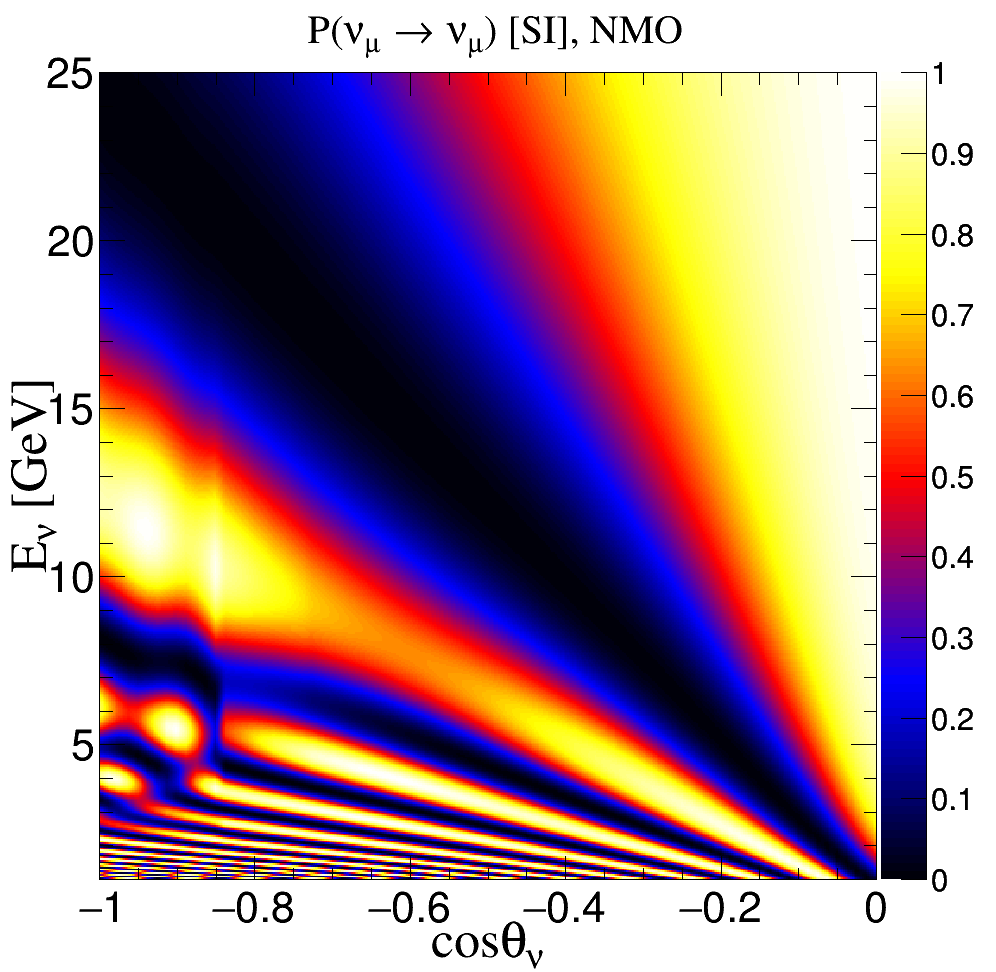}
  \hspace{0.1 cm}
  \includegraphics[width=0.425\textwidth]{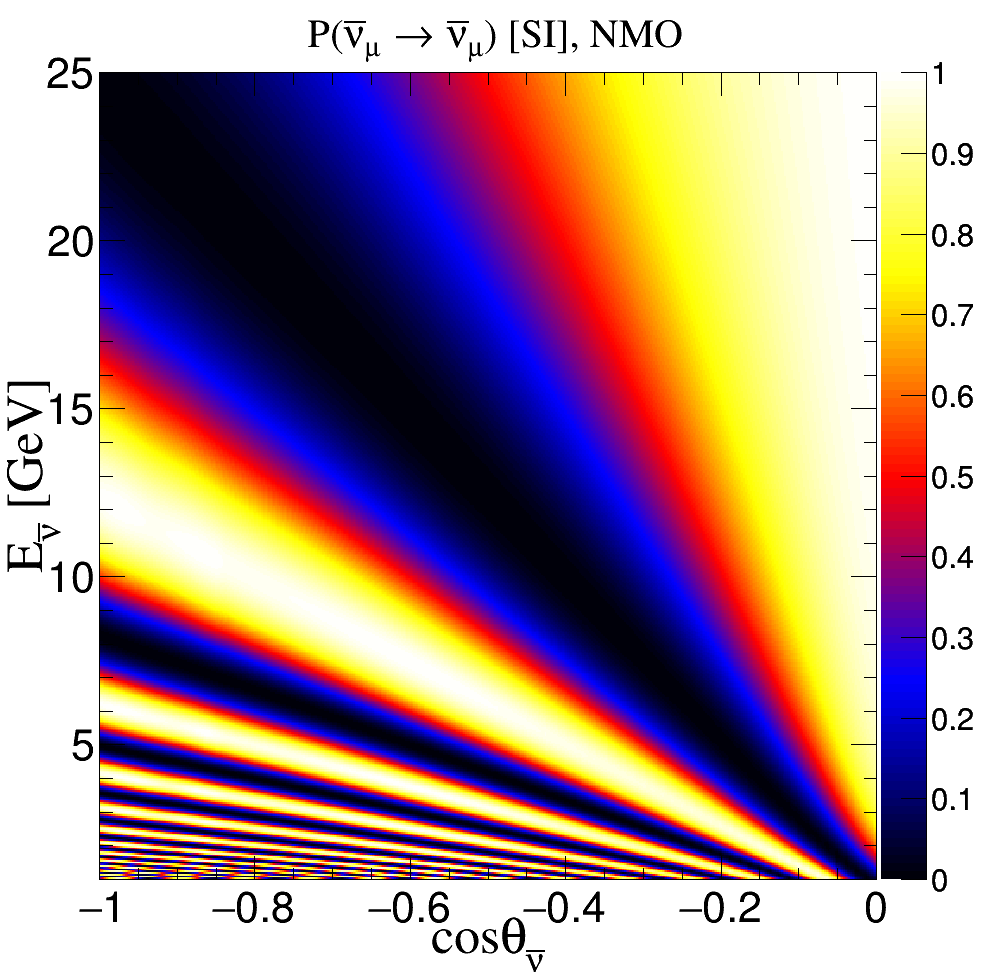}\\ 
  \vspace{0.1 cm}
  \includegraphics[width=0.425\textwidth]{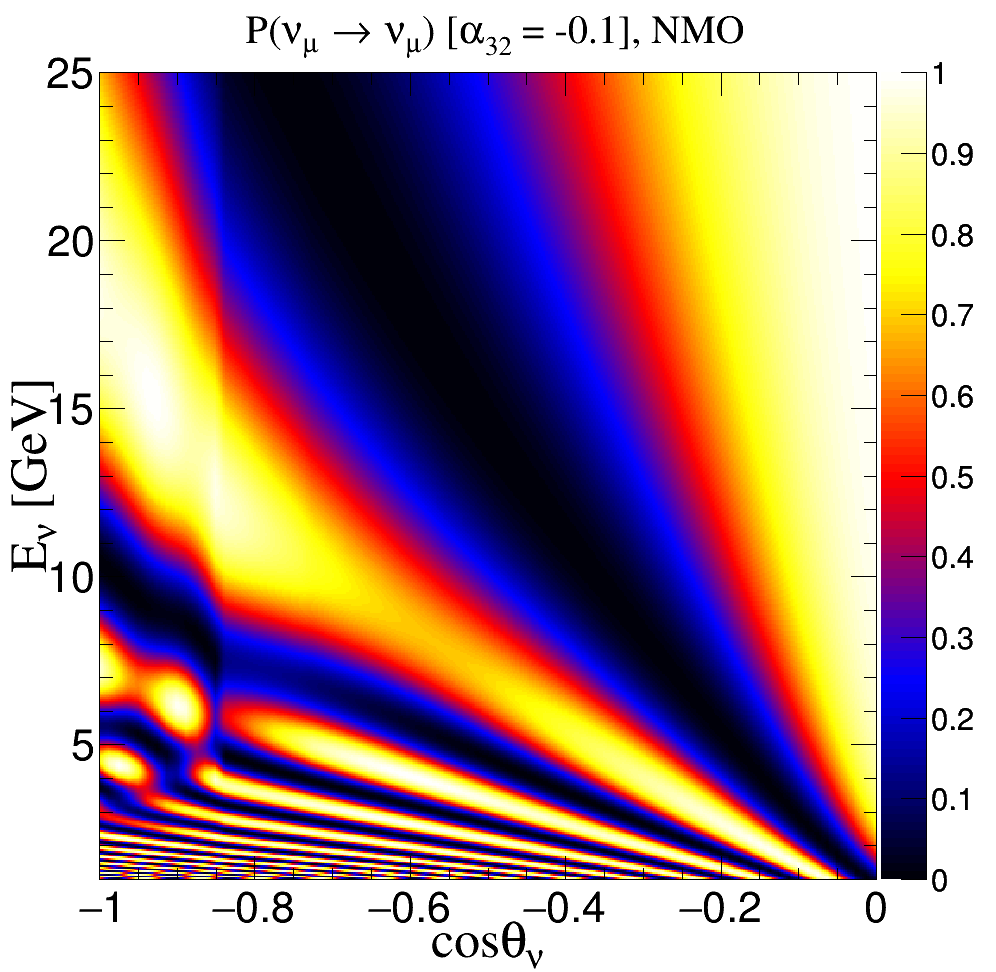}
  \hspace{0.1 cm}
  \includegraphics[width=0.425\textwidth]{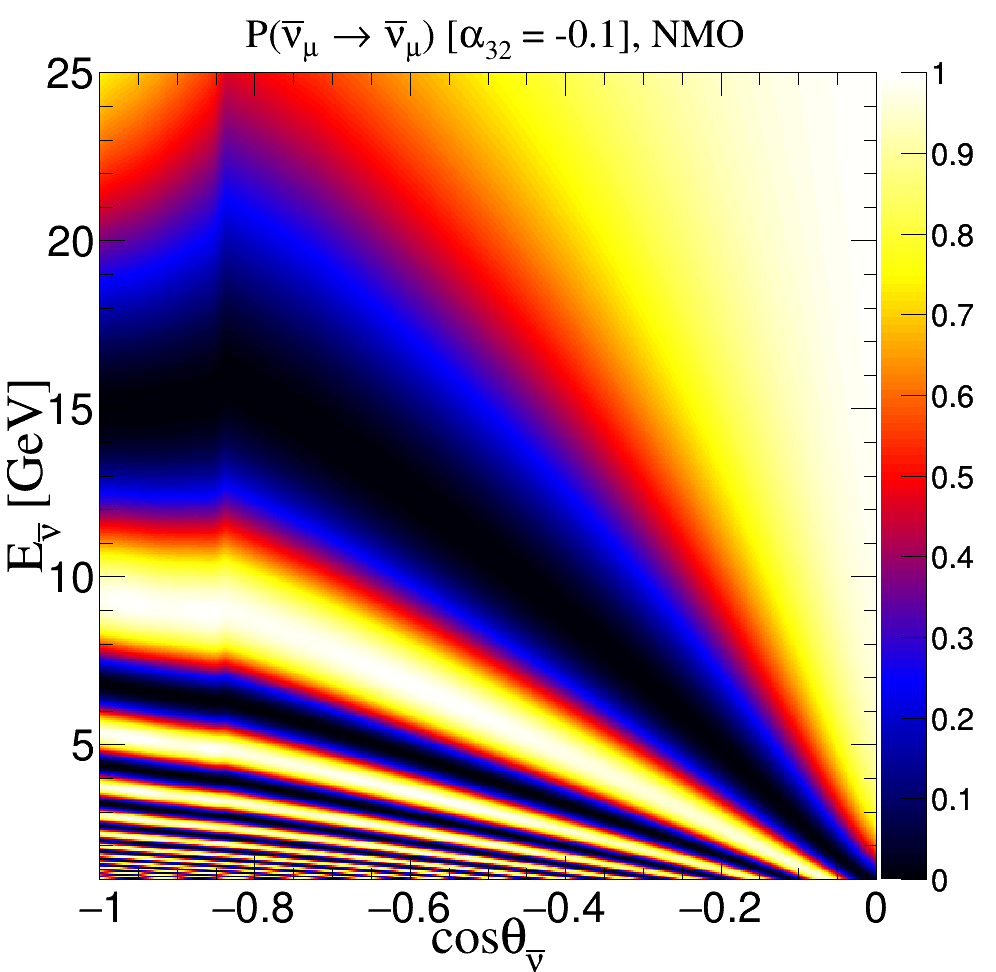}
  \mycaption{The survival probabilities of upward-going $\nu_\mu$ (left panels) and $\bar\nu_\mu$ (right panels) in the ($E_\nu$, $\cos\theta_\nu$) plane. We take three different values of the NUNM parameter $\alpha_{32}$ as $0.1$, $0$ (SI: Standard Interaction), and $-\,0.1$ in the top, middle, and bottom panels, respectively. We consider the Earth's matter effect assuming the PREM profile. The values of oscillation parameters are taken from table~\ref{Tab:1}.}
  \label{Fig:1}
\end{figure}
%=============================%

\begin{figure}[t]
  \centering
  \includegraphics[width=0.475\textwidth]{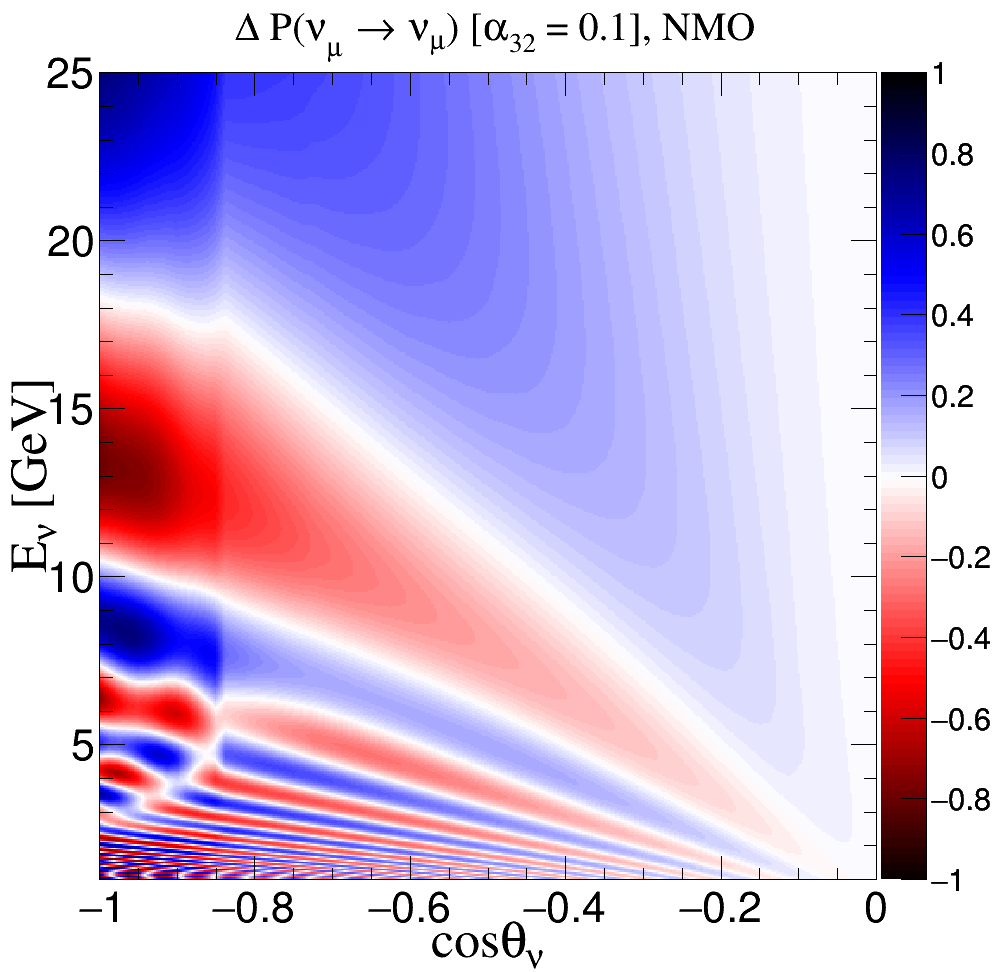}
  \hspace{0.25 cm}
  \includegraphics[width=0.475\textwidth]{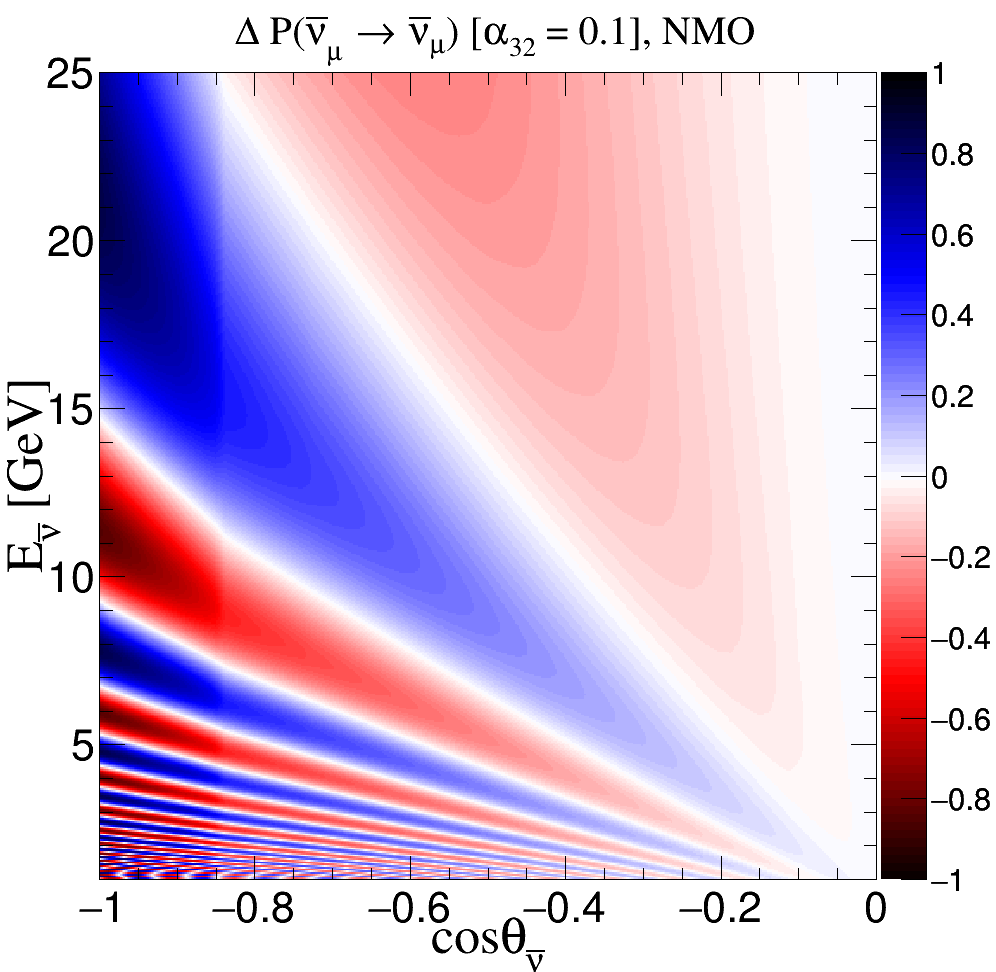}\\
  \vspace{0.25 cm}
  \includegraphics[width=0.475\textwidth]{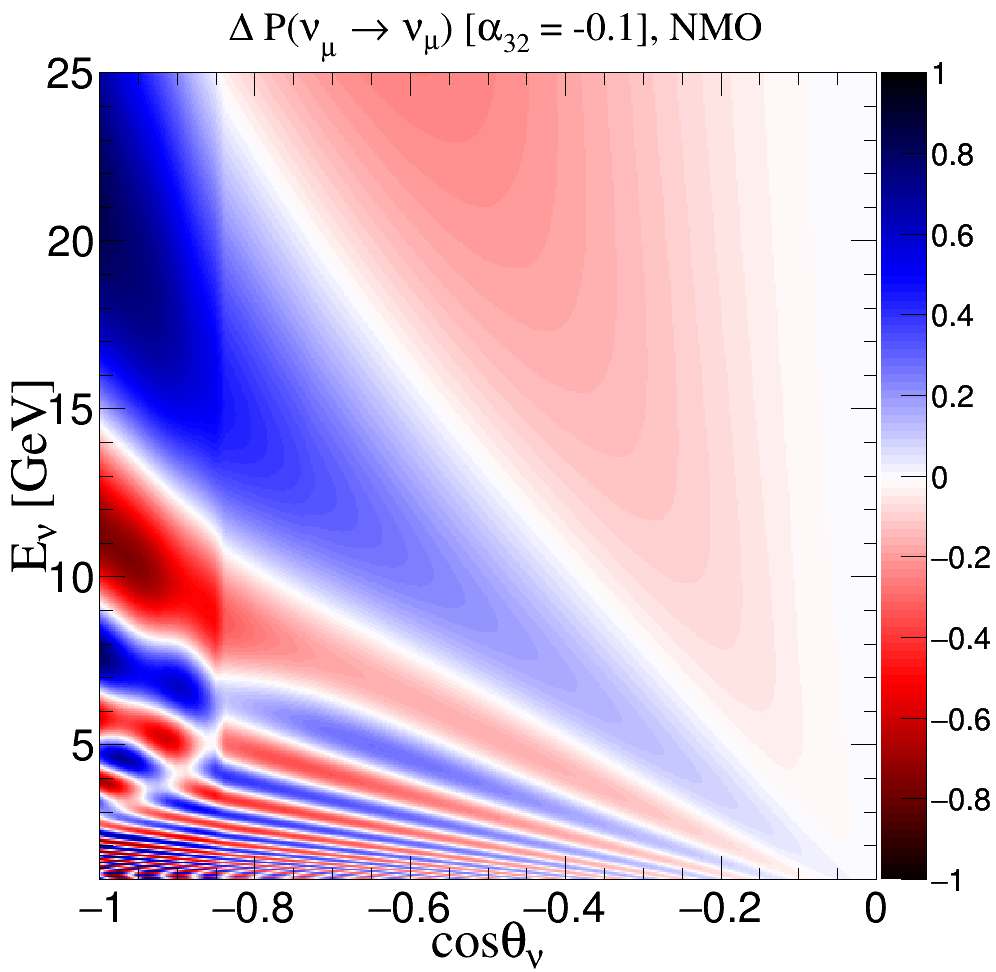}
  \hspace{0.25 cm}
  \includegraphics[width=0.475\textwidth]{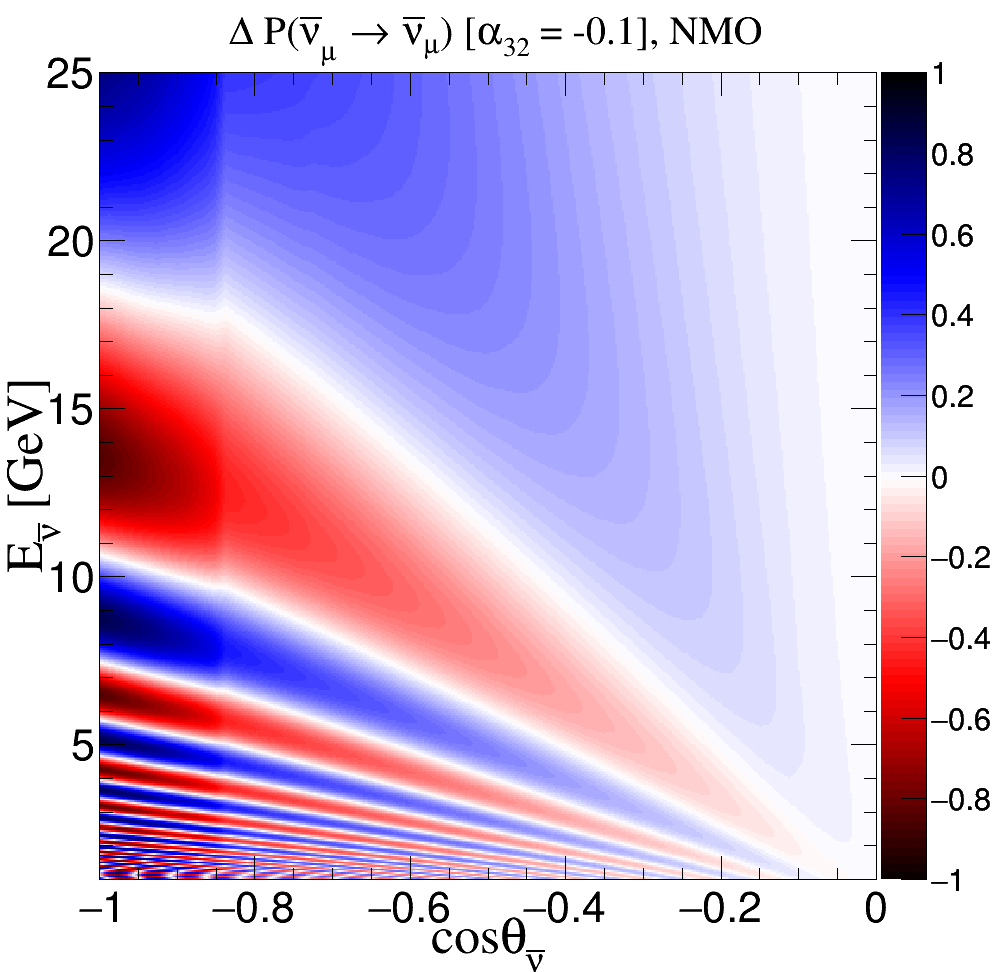}\\
  \mycaption{The survival probability difference $\Delta P$ (see ~(\ref{Eq:3.4})) for upward-going $\nu_\mu$ (left panels) and $\bar\nu_\mu$ (right panels) in the ($E_\nu$, $\cos\theta_\nu$) plane.  We take two different values of the NUNM parameter $\alpha_{32}$ as $0.1$ and $-\,0.1$ in the top and bottom panels, respectively. We consider the Earth's matter effect assuming the PREM profile. The values of oscillation parameters are taken from table~\ref{Tab:1}.}
  \label{Fig:2}
\end{figure}

In figure~\ref{Fig:1}, we demonstrate the impact of non-zero $\alpha_{32}$, with a representative value of 0.1, on the survival probability $P\left(\nu_\mu \to \nu_\mu\right)$ of upward-going multi-GeV $\nu_\mu$ and $\bar\nu_\mu$ in the plane of energy and zenith angle ($E_\nu,\;\cos\theta_\nu$). The left column represents $\nu_\mu$ survival probabilities for the values of $\alpha_{32} = 0.1$, $0$ (SI: Standard Interaction), and $-0.1$ in the top, middle, and bottom panels, respectively. The right column shows the same for $\bar\nu_\mu$. Note that the benchmark values of $\alpha_{32}=\pm\,0.1$ that we use in figures~\ref{Fig:1}-\ref{fig:4} and table~\ref{Event_Tab} are just for an illustration purpose. We deliberately use the values of $\alpha_{32}$ which are almost one order of magnitude higher than the current experimental limits. We do so to have a better visualization of the possible impact of 
 $\alpha_{32}$ on the oscillograms, event distributions, and while showing effective regions in ($E^{\rm rec}_\mu$, $\cos\theta^{\rm rec}_\mu$) plane to constrain $\alpha_{32}$.

Due to the effect of the NUNM parameter, the characteristic changes can be noticed in the oscillation valleys, which is a dark diagonal region with the least survival probability. It would be an almost triangular strip in a three neutrino unitary-mixing scenario in vacuum~\cite{Kumar:2020wgz, Kumar:2021lrn}. Here, we see a monotonic bending of these oscillation valleys in the range of $-0.85 < \cos\theta_\nu < 0$. Such a bending appears due to the term associated with $V_{\rm NC}$ in eq.~(\ref{Eq:2.6}) that offers an additional matter potential, i.e., due to neutral-current interaction along with the standard one. The monotonicity in the bending is directly related to the density profile of the Earth because in $-0.85 < \cos\theta_\nu < 0$, the Earth matter density has almost a monotonically increasing profile~\cite{Dziewonski:1981xy}. However, for $-1 \le \cos\theta_\nu \le -0.85$, neutrinos encounter the matter density at the Earth core region, which has a sharp rise in density profile. Thus, it involves a large matter interaction potential in eq.~(\ref{Eq:2.6}), which may introduce an extra orientation to the oscillation valleys at $-1 \le \cos\theta_\nu \le -0.85$. Since, for antineutrinos, both the matter potentials change their signs, the oscillation valleys bend in the opposite direction to that of neutrinos. To get an idea of how the NUNM parameter $\alpha_{32}$ impacts the survival probability $P(\nu_\mu \to \nu_\mu)$ and the oscillation valley, we contemplate an effective two neutrino scenario where $\alpha_{32}$ affects the $2$--$3$ sector of neutrino at the leading order.

Considering the one-mass-scale-dominance approximation, i.e., $\Delta m^2_{21}L/E \ll \Delta m^2_{32} L/E$, and $\sin^2\theta_{13} \simeq 0$, the survival probability $P(\nu_\mu \to \nu_\mu)$ can be expressed for a constant matter density and $\theta_{23} = 45^{\circ}$ as:\footnote{For a detailed derivation of $P(\nu_\mu \to \nu_\mu)$ in an effective two-flavor scenario with NUNM, see appendix~\ref{app:A2}.}
\begin{align}
  P\left(\nu_\mu \to \nu_\mu\right)\, =\, \cos^2\left[\bigg(\frac{\Delta m^2_{32}}{4E}\, +\, V_{NC} \cdot \alpha_{32}\bigg)\cdot L\right]\,,
  \label{Eq:3.1}
\end{align}
where $L$ is the path traversed through matter by neutrino of energy $E$. Using the condition of first oscillation minima in eq.~(\ref{Eq:3.1}), we can explain the feature of the bending of the oscillation valley due to the NUNM parameter $\alpha_{32}$. Assuming the propagation length of upward-going atmospheric neutrinos\footnote{The propagation length of atmospheric neutrinos traveling through the Earth can be calculated as: 
\begin{align}
  L\,=\, \sqrt{\left(R+h\right)^2\,-\,\left(R-d\right)^2\sin^2\theta_\nu}\, -\, \left(R-d\right)\cos\theta_\nu, 
  \label{Eq:3.2}
\end{align}
where $R$ is the radius of the Earth, and $h$ and $d$ stand for atmospheric height and detector depth from sea level, respectively. In our analysis, we consider $R = 6371$ km and $h = 15$ km while assuming that the detector (say ICAL) has been built at the sea level, i.e., $d =0$. Using $R\gg h$ approximation, the path length of upward-going neutrinos ($\cos\theta_\nu < 0$) can be approximated to be $L\simeq |2R\cos\theta_\nu|$.} to be $L\simeq |2R\cos\theta_\nu|$, and $Y_n \simeq 0.5$ with the value of $\rho$ using the line-averaged constant mass density approximation, the relation between $\cos\theta_\nu$ and $E$ can be written as \cite{Kumar:2021lrn}:
\begin{align}
  E\left[{\rm GeV}\right]\big\vert_{\rm valley} & \simeq \frac{1.03\times 10^{4}\cdot\Delta m^2_{32}\,[{\rm eV^2}]\cdot |\cos\theta_\nu|}{1\,\pm\,0.78\times\alpha_{32}\cdot\rho\,[{\rm g/cm^3}]\cdot|\cos\theta_\nu|}\,.
  \label{Eq:3.3}
\end{align}
Here, plus (+) sign stands for neutrino and minus (-) for antineutrino cases. For a given $\Delta m^2_{32}$, when $\alpha_{32}$ has null value (for SI case), $E \propto |\cos\theta_\nu|$ which indicates the valley at first oscillation minima to be a straight line. Now, considering the values of $\alpha_{32} > 0$ for neutrinos, the denominator becomes greater than that in the SI case, which implies lower neutrino energy $E$ and bending of oscillation valley in the downward direction. For $\alpha_{32} < 0$, the denominator decrease, and the oscillation valley bends in the upward direction. For antineutrino scenarios, the effects of non-zero $\alpha_{32}$ values on the $\cos\theta$ and $E$ relation are opposite to that of neutrinos. Using eq.~(\ref{Eq:3.3}), we get a qualitative idea of the impacts of $\alpha_{32}$ on the $\nu_\mu$ survival probabilities. However, to keep the analysis realistic throughout this paper, we use the three-flavor neutrino oscillation probabilities in the matter with the PREM profile. 
%=============================%
%=============================%

Due to access to a wide range of baselines and energies, atmospheric neutrino detectors are blessed with a unique advantage over fixed-baseline experiments. For instance, the neutrino can travel at most $\sim 1300$ km through a matter density profile of $\rho^{\rm max} < 3.0\;{\rm g/cm^3}$ for the longest fixed-baseline experiment (DUNE)~\cite{DUNE:2015lol, DUNE:2021cuw} to be built so far. On the other hand, a plethora of atmospheric neutrinos can travel $\sim 12750$ km through the Earth's Core region, where they can experience a maximum matter density of $\approx 13.5\;{\rm g/cm^3}$. Such a large matter density can alter the oscillation probability significantly as compared to the long-baseline one. To showcase this fact, we plot the difference between the survival probabilities of $\nu_\mu$ ($\bar\nu_\mu$) with NUNM and SI scenarios.
\begin{align}
  \Delta P = P_{\nu_\mu\to\nu_\mu}\left[\mathsmaller{{\rm NUNM}}\,(\alpha_{32}\ne 0)\right]\, - P_{\nu_\mu\to\nu_\mu}\left[\mathsmaller{{\rm SI}}\,(\alpha_{32} = 0)\right].
  \label{Eq:3.4}
\end{align}

In figure~\ref{Fig:2}, we demonstrate the probability difference $\Delta P$ in $\left(E,\,\cos\theta\right)$ plane, appearing due to non-zero values of NUNM parameter $\alpha_{32}$. The left (right) column shows the $\Delta P$ for upward-going $\nu_\mu$ ($\bar\nu_\mu$) with $\alpha_{32} = 0.1$ and $-\,0.1$ in the top and bottom panels, respectively. We find the value of $|\Delta P| \lesssim 0.45$ when neutrinos and antineutrinos traverse through the Earth's mantle region, i.e., for $\cos\theta_\nu \ge -0.85$. However, this probability difference gets amplified up to $0.83$ when they pass through the core of Earth, which is indeed reflected as the dark patches in the region $\cos\theta_\nu < -0.85$. Since these patterns are opposite for neutrinos and antineutrinos, the data from atmospheric neutrino detectors like ICAL, which is capable of observing $\nu_\mu$ and $\bar\nu_\mu$ events separately, can certainly test the prospects of minimal NUNM scenario.

%=============================%
\section{Simulation of neutrino events at the ICAL detector}
\label{sec:ICAL-detector}
%=============================%
To perform our analysis, we simulate neutrino events at the proposed 50 kt ICAL detector at INO~\cite{ICAL:2015stm}. ICAL would consist of three modules placed inside a cavern under the mountain in the Theni district of Tamilnadu, India. Each ICAL module of size 16 m $\times$ 16 m $\times$ 14.5 m contains 151 layers of magnetized iron plates of thickness $5.6$ cm stacked vertically with a gap of $4$ cm. These iron plates act as primary targets for the interactions of atmospheric $\nu_\mu$ and $\bar\nu_\mu$. The resistive plate chambers (RPCs)~\cite{Santonico:1981sc} of size $2$ m $\times$ $2$ m serve as active detectors and are sandwiched between two consecutive iron plates. The excellent timing resolution of RPC of $\sim1$ ns~\cite{Bheesette:2009yrp, Bhuyan:2012zzc, Bhatt:2019zsz}, helps in the discrimination of upward-going and downward-going moun events. In these RPCs, orthogonally installed pickup strips provide the coordinates of the event hits in the $X-Y$ plane, while the layer numbers of RPCs give the $Z$ coordinates.

ICAL can efficiently detect $\mu^-$ ($\mu^+$) events that have been produced during the charged-current interaction of multi-GeV $\nu_\mu$ ($\bar\nu_\mu$) via resonance and deep-inelastic scatterings. Since muon is the minimum ionizing particle in the multi-GeV range of energies, it leaves hits in the RPCs in the form of a long track. ICAL can measure the reconstructed muon energy in the range of 1 to 25 GeV with a resolution of about 10 to 15\%~\cite{Chatterjee:2014vta}. The excellent resolution of the reconstructed muon zenith angle of $< 1^\circ$ helps to infer the path length traversed by the parent neutrinos~\cite{Chatterjee:2014vta}. The deep-inelastic scatterings of muon neutrinos can also produce hadron showers, which carry away a fraction of incoming neutrino energy. The energy deposited in the form of hadron shower can be given as $E^\prime_{\rm had} \equiv E_\nu - E_\mu$, where $E_\nu$ and $E_\mu$ are the true energies of incoming neutrino and generated muon, respectively.

Along with the energy and direction of the reconstructed muon, ICAL can measure the energy of the hadron shower with a resolution of about 35 to 70\%~\cite{Devi:2013wxa} on an event-by-event basis. The applied magnetic field of $\sim1.5$ Tesla in ICAL gives a unique feature of the charge identification (CID) capability to observe $\mu^-$ and $\mu^+$ events separately~\cite{Behera:2014zca}. For muons of energy of a few GeV to $50$ GeV, the CID efficiency of ICAL lies between $98$ to $99$\%, which helps in distinguishing the interactions of atmospheric $\nu_\mu$ and $\bar\nu_\mu$ events efficiently.

In this work, we simulate the unoscillated neutrino events generated via the charged-current interactions using the NUANCE Monte Carlo neutrino event generator~\cite{Casper:2002sd}. During this simulation, we consider the geometry of ICAL as target and  use the 3D Honda flux~\cite{SajjadAthar:2012dji,Honda:2015fha} of atmospheric neutrinos estimated for the proposed INO site at Theni district of Tamil Nadu, India as a source. Here, we consider the effects of solar modulation on atmospheric neutrino flux by including solar maxima for half of its exposure and solar minima for another half. A rock coverage of at least $\sim 1$ km from above would help in reducing the downward-going cosmic muon background at ICAL by a factor of $\sim 10^6$~\cite{Dash:2015blu}. The muon events entering into the detector from outside are excluded by considering only those events in the analysis that have vertices inside the detector and far from the edges. Therefore, we expect negligible background at ICAL due to the downward-going cosmic muons. We do not consider the muon events produced during the decay of tau leptons originated from $\nu_\tau$ interactions because these lower energy events are expected to be small (only $\sim$2\% of the total up-going muon events from $\nu_\mu$ interactions~\cite{Pal:2014tre} ) and mostly below the 1 GeV energy threshold of ICAL.
%=============================%
\begin{table}[t]
  \centering
  \begin{tabular}{|c|c|c|c|c|}
    \hline
    Event type & Oscillation channels & $\alpha_{32} = 0$ [SI] & $\alpha_{32} = -\,0.1$ & $\alpha_{32} = \,0.1$\\
    \hline
    \hline
    \multirow{3}{*}{$\mu^-$} & ${\nu_\mu \to \nu_\mu}$ & 4318 & 4342 & 4329 \\
    \cline{2-5}
    &${\nu_e \to \nu_\mu}$        & 95      & 99     & 92       \\ 
    \cline{2-5}
    & $({\nu_\mu \to \nu_\mu})\;+\;({\nu_e \to \nu_\mu})$&  4413 & 4441 & 4421 \\
    \hline
    \hline
    \multirow{3}{*}{$\mu^+$} &${\bar\nu_\mu \to \bar\nu_\mu}$ & 2002 & 2003 & 2011 \\
    \cline{2-5}
    & ${\bar\nu_e \to \bar\nu_\mu}$ & 12 & 11 & 12 \\
    \cline{2-5}
    & ${(\bar\nu_\mu \to \bar\nu_\mu})\;+\;({\bar\nu_e \to \bar\nu_\mu})$ & 2014 & 2014 & 2023 \\
    \hline
  \end{tabular}
  \mycaption{The expected number of reconstructed $\mu^-$ ($\mu^+$) events obtained from $500$ kt$\cdot$yr exposure of ICAL via both survival and appearance channels of atmospheric $\nu_\mu$ ($\bar\nu_\mu$). We show total $\mu^-$ (top part) and $\mu^+$ (bottom part) event rates considering $\alpha_{32} = 0$ [SI: Standard Interaction] (third column), $-\,0.1$ (fourth column), and $0.1$ (fifth column). We use the values of oscillation parameters as given in table~\ref{Tab:1}.}
\label{Event_Tab}
\end{table}
%=============================%

Since we estimate median (Asimov) sensitivities in the present work, the statistical fluctuations are needed to be suppressed. This is achieved by simulating the events for a large exposure of $50$ Mt$\cdot$yr which is then scaled down to $500$ kt$\cdot$yr for sensitivity calculations. We employ the reweighting algorithm given in refs.~\cite{Ghosh:2012px, Thakore:2013xqa} to incorporate the three-flavor neutrino oscillations for the NUNM scenario in the presence of Earth's matter effects with PREM profile~\cite{Dziewonski:1981xy}.
%=============================%
\begin{table}[b] 
  \centering 
  \begin{tabular}{|c| c| c| c|} 
    \hline 
    Observable & Range & Bin width & Total bins \\ 
    \hline
    $E_{\mu}^\text{rec}$ (GeV) & \makecell[c]{$[1, 11]$ \\ $[11, 21]$ \\ $[21, 25]$} & 
    \makecell[c]{1 \\ 5 \\ 4} & 
    $\left.\begin{tabular}{l}
      10 \\ 2 \\ 1
    \end{tabular}\right\}$  13 \\
    \hline
    $\cos\theta_\mu^\text{rec}$  & \makecell[c]{$[-1.0, 0.0]$ \\ $[0.0, 1.0]$} & 
    \makecell[c]{0.1 \\ 0.2} & 
    $\left.\begin{tabular}{l}
      10 \\ 5
    \end{tabular}\right\}$  15 \\
    \hline
    ${E'}_\text{had}^\text{rec}$ (GeV)  & \makecell[c]{$[0, 2]$ \\ $[2, 4]$ \\ $[4, 25]$} &
    \makecell[c]{1 \\ 2 \\ 21} & 
    $\left.\begin{tabular}{l}
      2 \\ 1 \\ 1
    \end{tabular}\right\}$  4\\ \hline
  \end{tabular}
  \mycaption{The binning scheme adopted in this analysis for the three reconstructed observables $E_\mu^\text{rec}$, $\cos\theta_\mu^\text{rec}$, and ${E'}_\text{had}^\text{rec}$. We use the same binning scheme for reconstructed $\mu^-$ and $\mu^+$ events as done in ref.~\cite{Sahoo:2021dit}.}
  \label{Tab:2}
\end{table}
%=============================%
\noindent
We incorporate the detector response in terms of reconstruction efficiency, CID efficiency, and resolutions of muon energy, muon zenith angle and hadron energy, using the ICAL migration matrices~\cite{Chatterjee:2014vta, Devi:2013wxa} that has been approved by the ICAL Collaboration. The reconstructed $\mu^-$ and $\mu^+$ events would have observables such as muon energy $(E_{\mu}^{\rm rec})$, zenith angle $(\cos\theta_\mu^{\rm rec})$, and hadron energy $({E'}_{\rm had}^{\rm rec})$.

For an exposure of 500 kt$\cdot$yr, we expect about 4400 $\mu^-$ and 2000 $\mu^+$ reconstructed events at ICAL. In table~\ref{Event_Tab}, we show the expected contributions to total reconstructed $\mu^-$ events via $\nu_\mu \to \nu_\mu$ survival and $\nu_e \to \nu_\mu $ appearance channels for the NUNM scenarios ($\alpha_{32} = -\,0.1$ and $0.1$), and compare with standard scenario ($\alpha_{32} = 0$). We also show a similar comparison for the total reconstructed $\mu^+$ events. Even though the difference in the total number of events due to the presence of non-zero $\alpha_{32}$ may not be large, a significant modification can be observed in the binned distribution of reconstructed events in $E_{\mu}^{\rm rec}$, $\cos\theta_\mu^{\rm rec}$, and ${E'}_{\rm had}^{\rm rec}$.
%=============================%
\begin{figure}[t]
  \centering
  \includegraphics[width=0.32\textwidth]{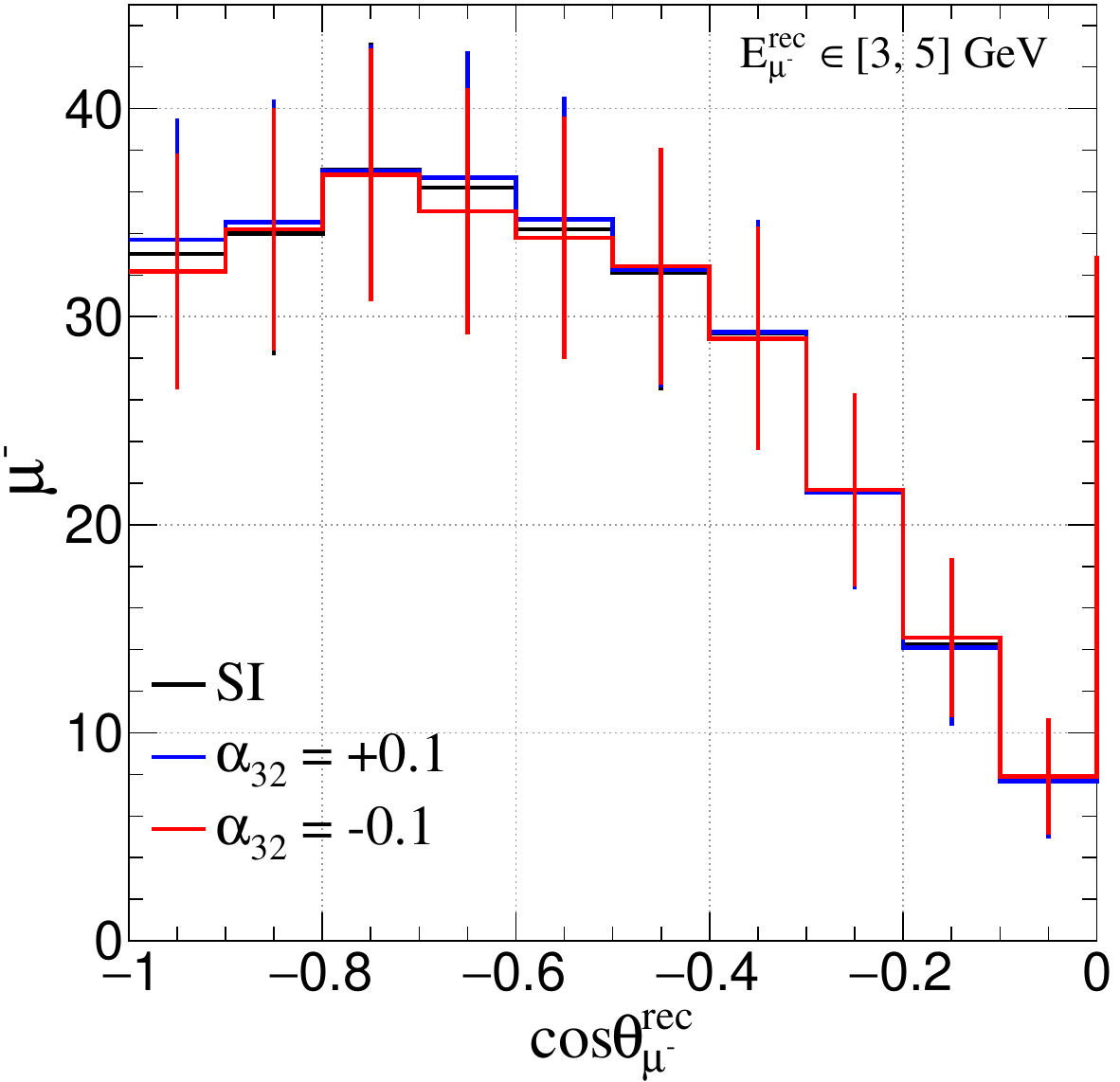}
  \includegraphics[width=0.32\textwidth]{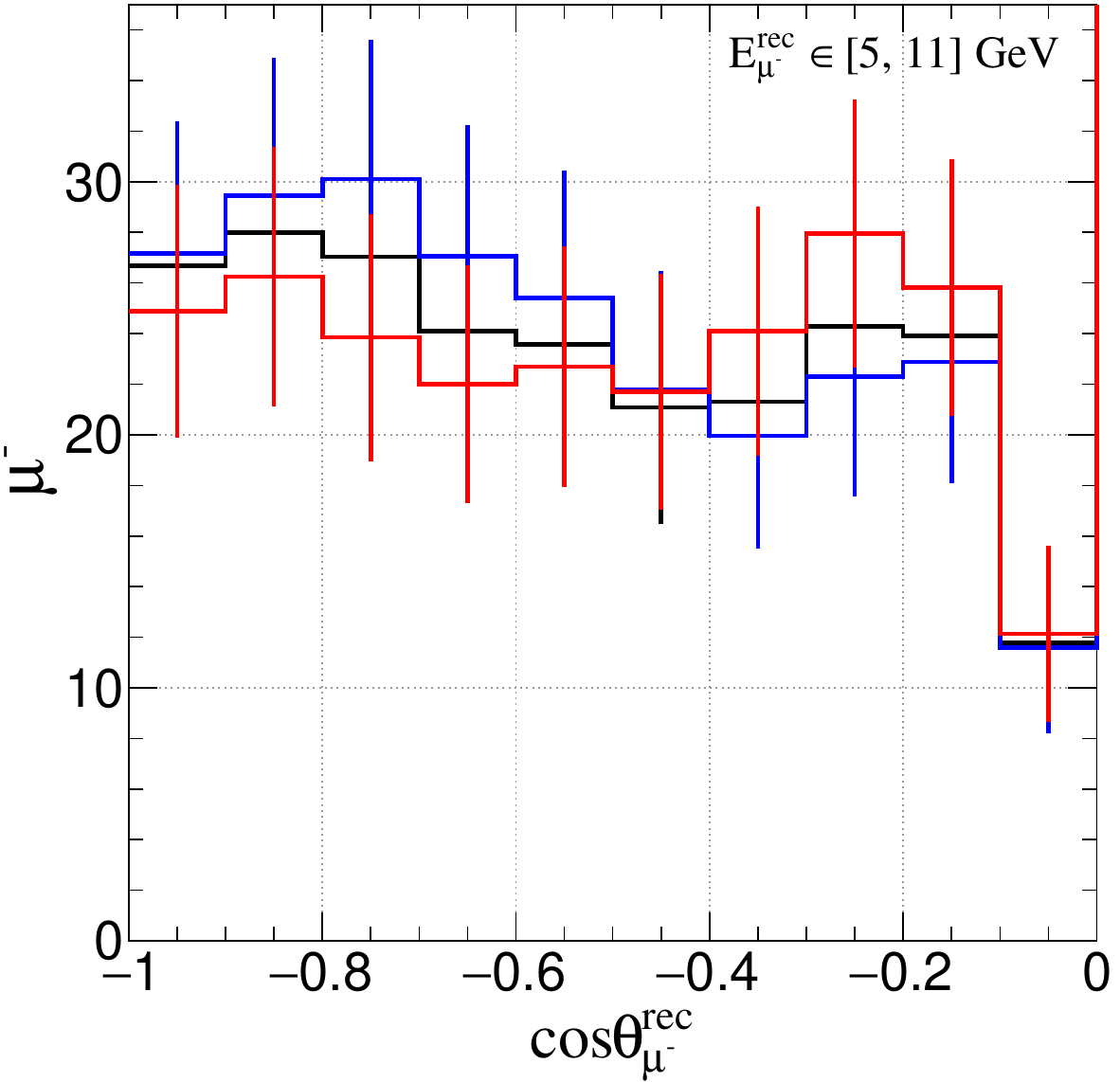}
  \includegraphics[width=0.32\textwidth]{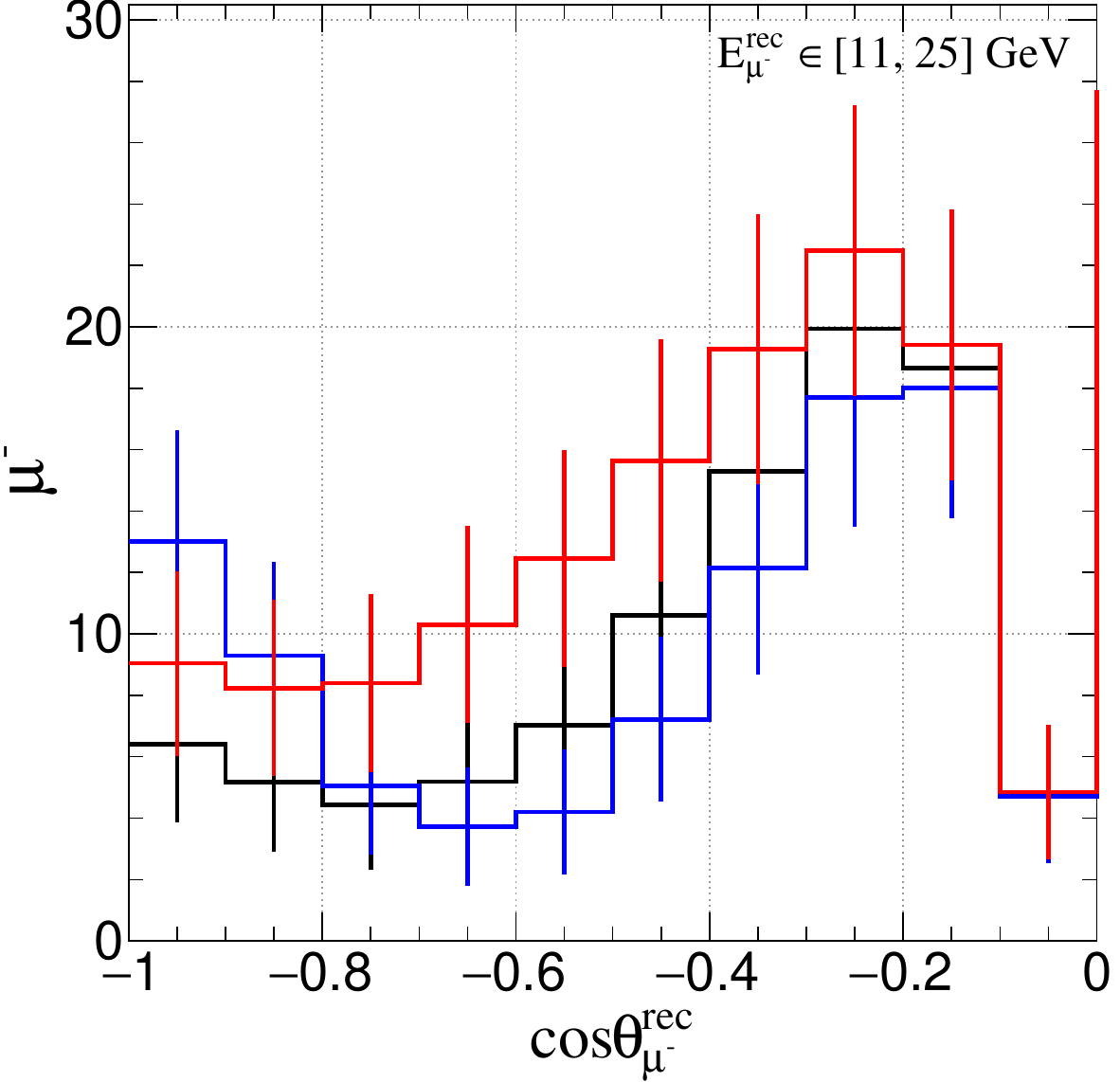}\\
  \includegraphics[width=0.32\textwidth]{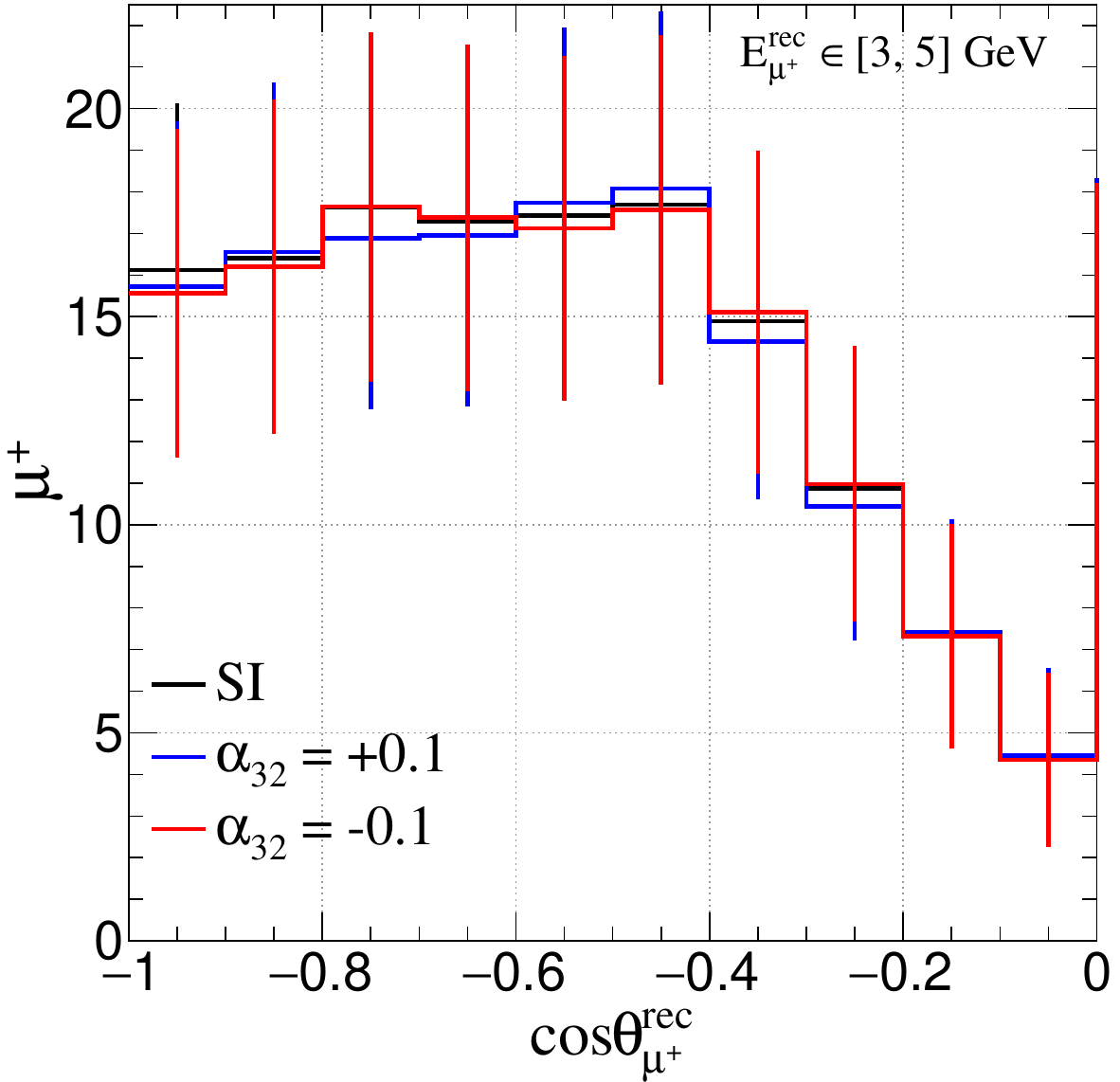}
  \includegraphics[width=0.32\textwidth]{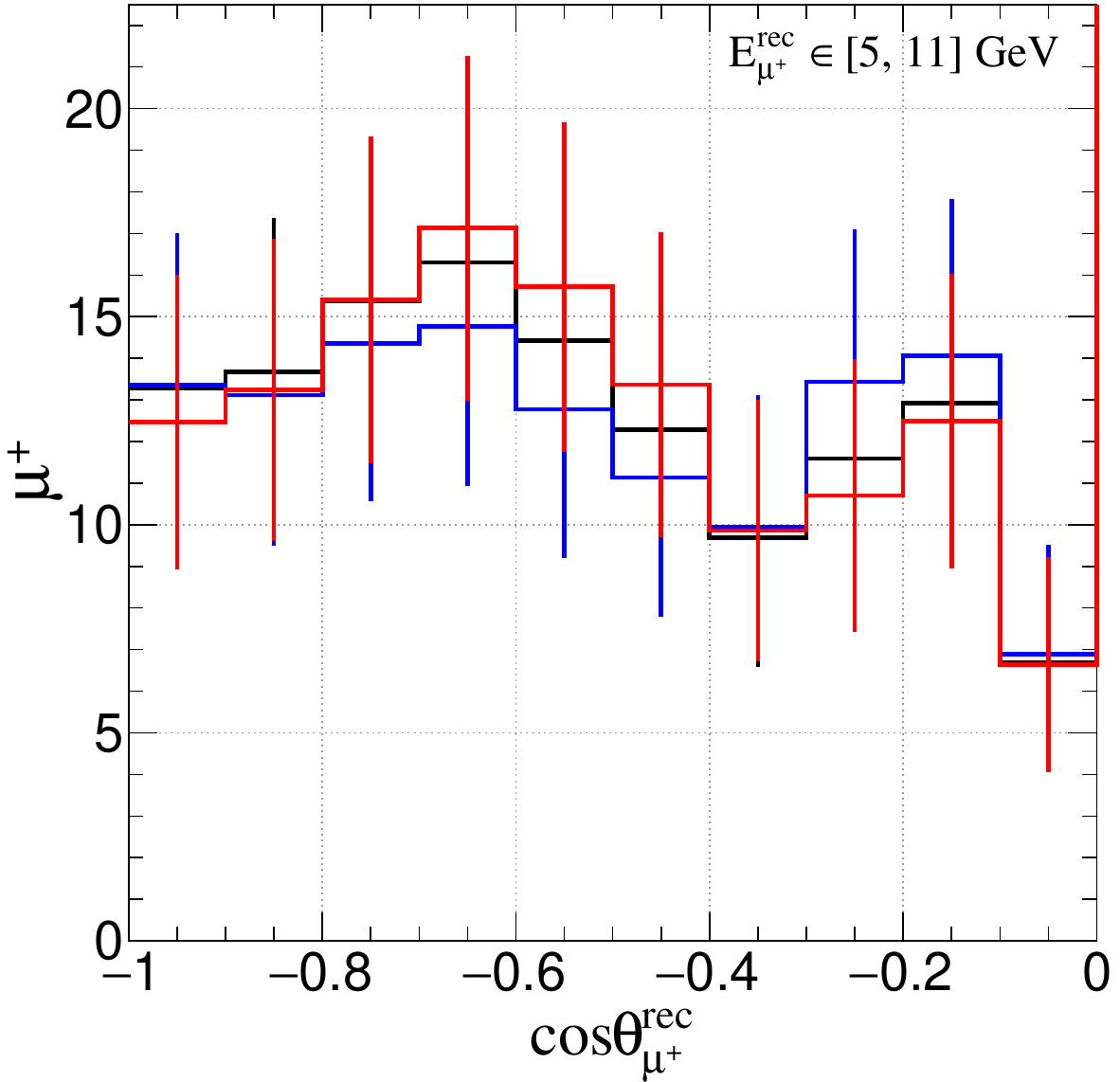}
  \includegraphics[width=0.32\textwidth]{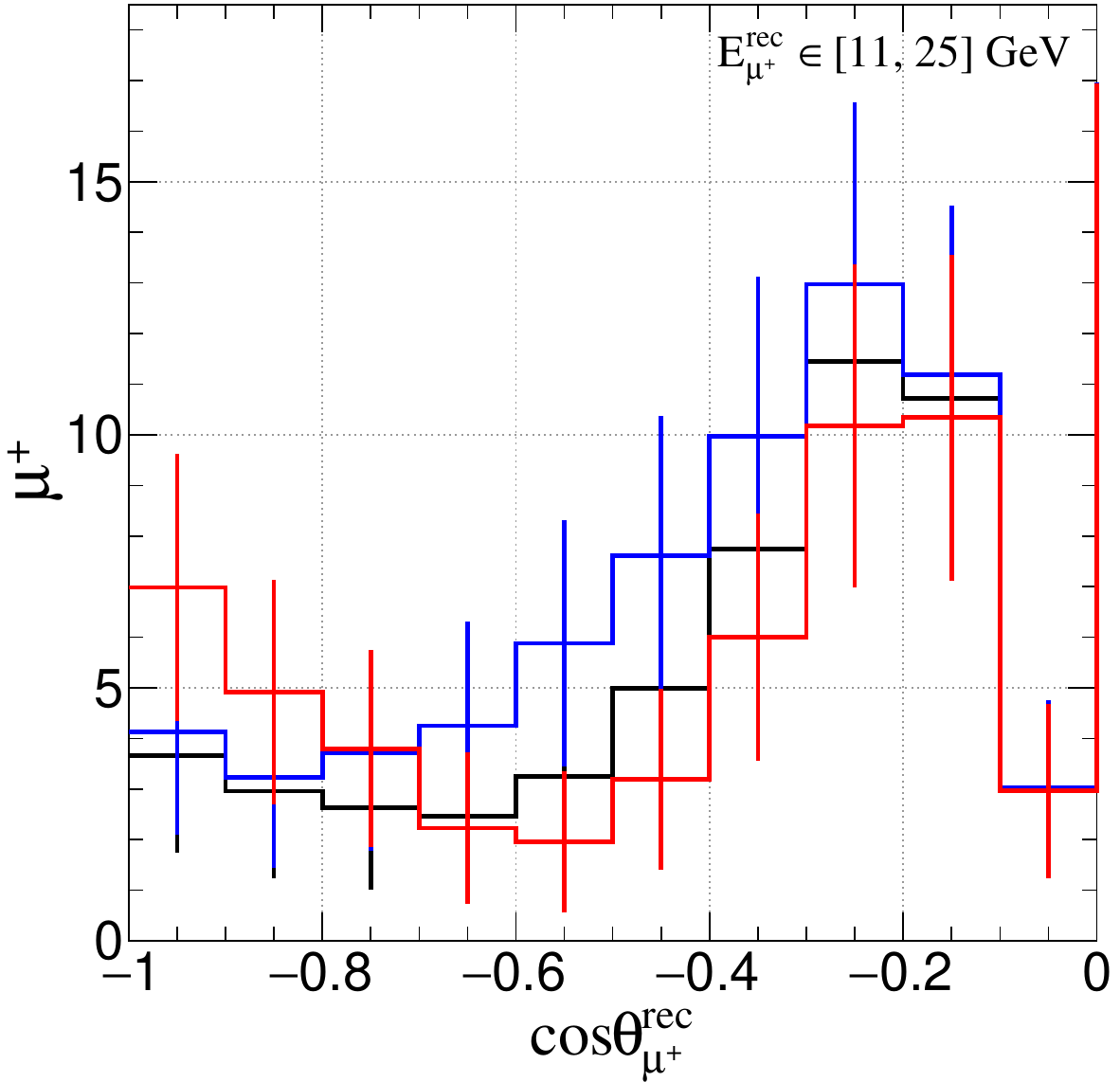}
  \mycaption{The upward-going reconstructed $\mu^-$ ($\mu^+$) event distributions are shown in the top (bottom) panels for $500$ kt$\cdot$yr exposure of ICAL. The black, blue, and red lines in the plots correspond to SI ($\alpha_{32} = 0$), $\alpha_{32} = 0.1$, and $\alpha_{32} = -\,0.1$, respectively. We consider three different ranges of $E_\mu^{\rm rec}$: $\left[3,\, 5\right]$ GeV (left panels), $\left[5,\, 11\right]$ GeV (middle panels), and $\left[11,\, 25\right]$ GeV (right panels). Note that the range of the y-axis is different in each plot. We use the values of oscillation parameters given in table~\ref{Tab:1}, and the binning scheme specified in table~\ref{Tab:2}, where the recontructed hadron energy variable ${E^\prime}_{\rm had}^{\rm rec}$ is integrated over the range of $0$ to $25$ GeV. }
\label{Fig:3}
\end{figure}
%=============================%

In figure~\ref{Fig:3}, we show the distributions of upward-going reconstructed $\mu^-$ (top row) and $\mu^+$ (bottom row) events for $500$ kt$\cdot$yr exposure at ICAL. We consider the values of oscillation parameters mentioned in table~\ref{Tab:1} while incorporating three-flavor neutrino oscillations in the presence of matter. To plot the distributions, we use the binning scheme given in table~\ref{Tab:2}. Here, we show the event distributions for three different ranges of reconstruction muon energies $E_{\mu}^{\rm rec}$ which are $\left[3,\, 5\right]$ GeV (left panel), $\left[5,\, 11\right]$ GeV (mid panel), and $\left[11,\, 25\right]$ GeV (right panel) while integrating reconstructed hadron energies $\left({E'}_{\rm had}^{\rm rec}\right)$ over $0$ to $25$ GeV. The black, blue, and red curves correspond to three different physics scenarios, i.e., $\alpha_{32} = 0$ (SI), $0.1$, and $-0.1$, respectively. The error bars represent the statistical fluctuations. The impacts of non-zero NUNM parameters $\left(\alpha_{32} = \pm\, 0.1\right)$ can be visually appreciated when $E_{\mu}^{\rm rec} > 5$ GeV, for both reconstructed $\mu^-$ and $\mu^+$ events. Now, it is essential to note that for a given non-zero value of $\alpha_{32}$, the event distributions of $\mu^-$ and $\mu^+$ have opposite orientations with respect to SI. For an instance, the bin contents of blue curve ($\alpha_{32} = +0.1)$ for $\mu^-$ distributions in reconstructed muon energy range [11, 25] GeV have a lower value than the SI ones, whereas these get higher for the $\mu^+$ distributions. Such distinguishing characteristics of new physics scenarios can only be observed efficiently when the detector has the capability to identify neutrino and antineutrino events separately, like at ICAL; otherwise, it would be diluted.  

%=============================%
\section{Statistical analysis}
\label{sec:Stat_Meth}
%=============================%
In this analysis, we use a frequentist approach to get the median sensitivity of the detector while testing the hypothesis of non-unitary neutrino mixing. Here, we calculate the Poissonian $\chi^2$~\cite{Blennow:2013oma} for the reconstructed $\mu^-$ and $\mu^+$ events by minimizing it over the systematic uncertainties. We define the $\chi^2$ while considering three reconstructed observables: $E_\mu^{\rm rec}$, $\cos\theta_\mu^{\rm rec}$ and ${E'}_{\rm had}^{\rm rec}$ as follows~\cite{Devi:2014yaa}: 
\begin{equation}
  \chi^2\left(\mu^\pm\right) = \mathop{\text{min}}_{\xi_l} \sum_{i=1}^{N_{{E'}_\text{had}^\text{rec}}} \sum_{j=1}^{N_{E_{\mu^\pm}^\text{rec}}} \sum_{k=1}^{N_{\cos\theta_{\mu^\pm}^\text{rec}}} 2\left[(N_{ijk}^\text{thr} - N_{ijk}^\text{obs})\,+\,N_{ijk}^\text{obs} \ln\left(\frac{N_{ijk}^\text{obs} }{N_{ijk}^\text{thr}}\right)\right] + \sum_{l = 1}^5 \xi_l^2\,,
  \label{Eq:5.1}
\end{equation}
where 
\begin{equation}
  N_{ijk}^\text{thr} = N_{ijk}^{0\; \text{thr}}\left(1 + \sum_{l=1}^5 \pi^l_{ijk}\xi_l\right).
  \label{Eq:5.2}
\end{equation}
Here, $N_{ijk}^\text{thr}$ stands for the theoretically expected number of reconstructed muon events while $N_{ijk}^\text{obs}$ stands for observed muon events, in a given bin of $(E_\mu^\text{rec}, \cos\theta_\mu^\text{rec}, {E'}_\text{had}^\text{rec})$. $N_{ijk}^{0\; \text{thr}}$ represents the pure theoretical prediction of reconstructed events. However, the systematic uncertainties ($\pi^l_{ijk}$) can modify the pure predicted events. Thus, we adopt the pull method~\cite{GonzalezGarcia:2004wg,Huber:2002mx,Fogli:2002pt} to address such fluctuations. Using the pull method, we parameterize the systematic uncertainties in terms of a set of variables called pull variables $\xi_l$. Here, we consider a linearized approximation while accounting for the five systematic uncertainties as 20\% flux normalization error, 10\% error in cross section, 5\% energy dependent tilt error in flux, 5\% uncertainty on the zenith angle dependence of the flux, and 5\% overall systematics for both $\nu_\mu$ and $\bar\nu_\mu$ events, as prescribed in the refs.~\cite{Kameda:2002fx, Devi:2014yaa}. The total $\chi^2$ is a sum of $\chi^2\left(\mu^-\right)$ and $\chi^2\left(\mu^+\right)$ as:
\begin{equation}
  \chi^2 = \chi^2\left(\mu^-\right) \, + \, \chi^2\left(\mu^+\right).
  \label{Eq:5.3}
\end{equation}
Now, we define the new physics sensitivity of ICAL in terms of $\Delta \chi^2$ as:
\begin{align}
  \Delta \chi^2 & = \chi^2\big(\rm NUNM\big)\, - \,\chi^2\big(\rm SI\big),
  \label{Eq:5.4}
\end{align}
where $\chi^2\big(\rm NUNM\big)$ and $\chi^2\big(\rm SI\big)$ are obtained by fitting the MC data with NUNM and SI scenarios, respectively. The MC data is generated assuming the SI scenario with the values of oscillation parameters as given in table~\ref{Tab:1} while considering NMO. In theory, we keep all the three diagonal NUNM parameters $\alpha_{ii}$, and the two off-diagonal NUNM parameters ($\alpha_{21},\;\alpha_{31}$) as zero, while considering only the non-zero real values of $\alpha_{32}$. Since the statistical fluctuations are suppressed for calculating the median sensitivity of ICAL, we have $\chi^2\big(\rm SI\big) \approx 0$. Note that the ICAL data is sensitive to $\sin^2\theta_{23}$, $\Delta m^2_{32}$, and the neutrino mass ordering. Therefore, in this paper, we minimize the $\Delta\chi^2$ in the fit over $\sin^2\theta_{23}$ in the range $0.36$ to $0.66$, $\Delta m^2_{32}$ in the interval $\left[2.1,\,2.6\right]\times 10^{-3}\;{\rm eV^2}$ for NMO in order to estimate the possible constraint on the NUNM parameter $\alpha_{32}$. Note that we also minimize the $\Delta\chi^2$ over both the choices of mass ordering, i.e., NMO and IMO. The recipe to switch the values of $\Delta m^2_{32}$ from NMO to IMO via $\Delta m^2_{\rm eff}$ is already discussed in section~\ref{sec:atm-tool}. 
%=============================%
\section{Results}
\label{sec:Results}
%=============================%
Before we start discussing our sensitivity results on the NUNM parameter $\alpha_{32}$, it is important to study the effective ranges of reconstructed muon variables $\left(E^{\rm rec}_\mu,\;\cos\theta^{\rm rec}_\mu\right)$ which contribute to our sensitivity. The following subsection is devoted to shed light on this issue.
%=============================%
\subsection{Effective regions in $\left(E^{\rm rec}_\mu,\;\cos\theta^{\rm rec}_\mu\right)$ plane to constrain $\alpha_{32}$}
%=============================%
%=============================%
\begin{figure}[ht]
  \centering
  \includegraphics[width=0.475\textwidth]{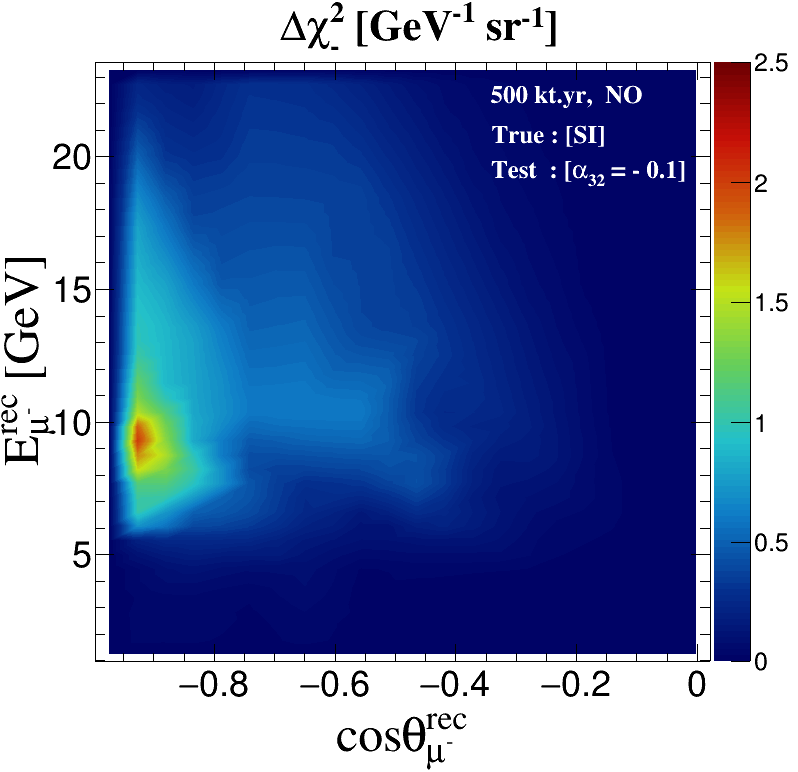}
  \includegraphics[width=0.475\textwidth]{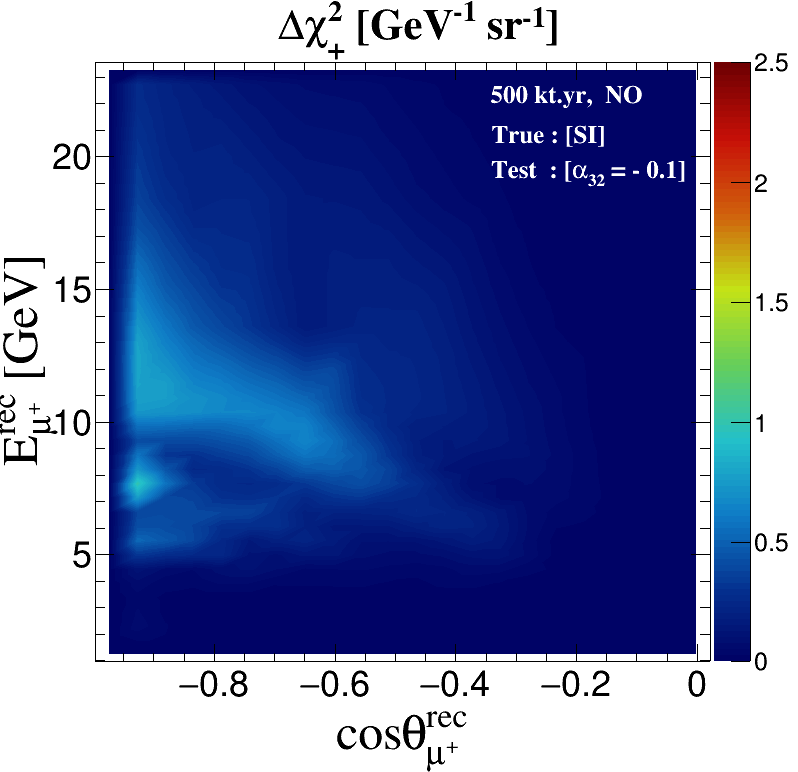}
  \includegraphics[width=0.475\textwidth]{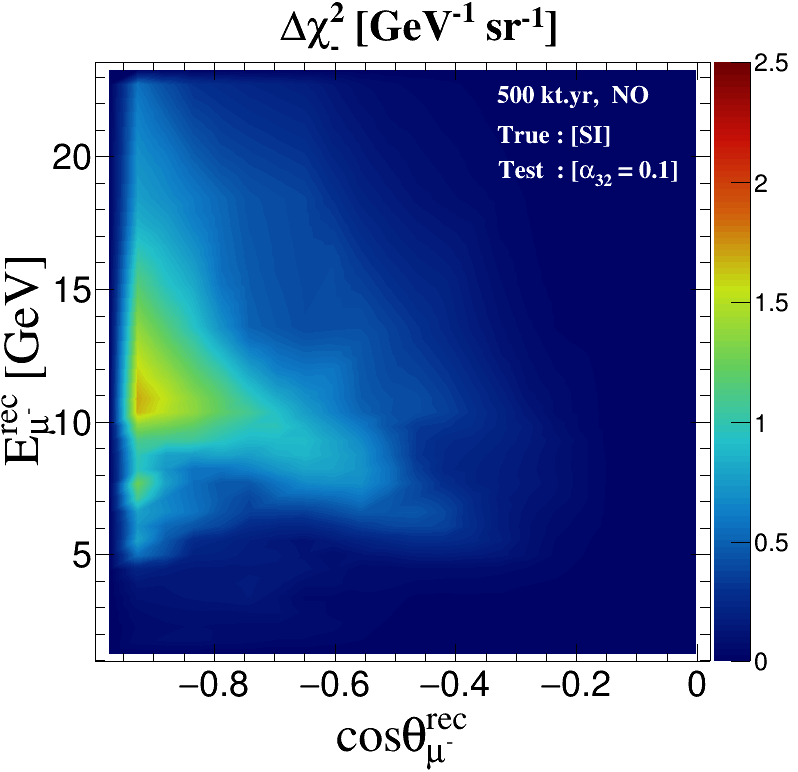}
  \includegraphics[width=0.475\textwidth]{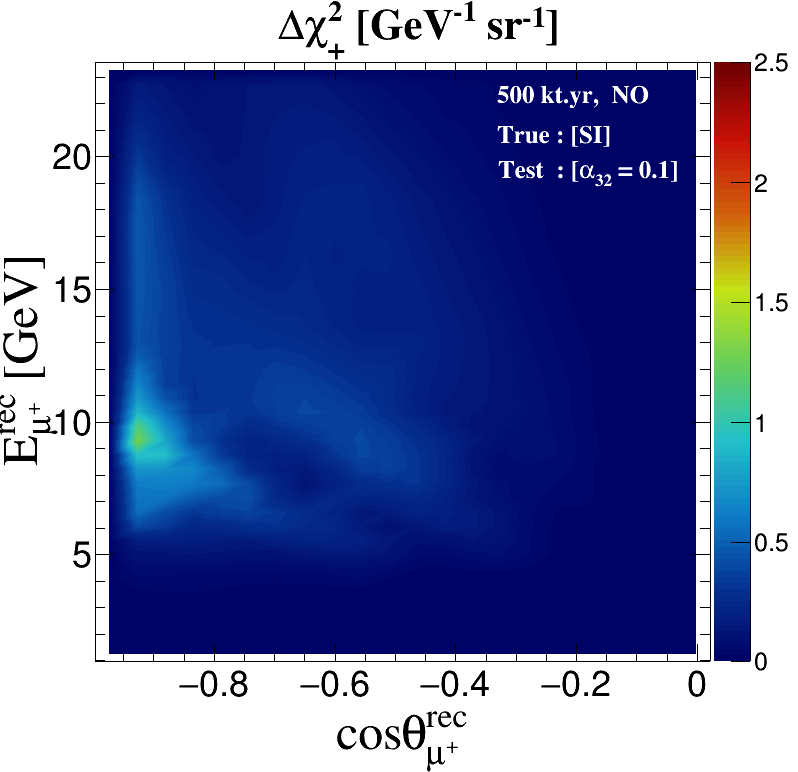}
  \mycaption{We show the distributions of fixed-parameter $\Delta\chi^2_-$ ($\Delta\chi^2_+$) in the plane of ($E_\mu^\text{rec}$, $\cos\theta_\mu^\text{rec}$) without the pull penalty term in the left (right) panels using 500 kt$\cdot$yr exposure of the ICAL detector. Note that $\Delta\chi^2_-$ and $\Delta\chi^2_+$ are presented in the units of $\text{GeV}^{-1} \text{sr}^{-1}$ where we have divided them by 2$\pi ~ \times$ bin area. The prospective data is generated assuming $\alpha_{32} = 0$, NMO (true), and using the benchmark values of oscillation parameters given in table~\ref{Tab:1}. In theory, $\alpha_{32} = -\,0.1$ and $0.1$ in the top and bottom panels, respectively.}
  \label{fig:4}
\end{figure}
%=============================%
We estimate the density distributions of $\Delta\chi^2$ $\left[\rm GeV^{-1}\, sr^{-1}\right]$ in the plane of $\left(E^{\rm rec}_\mu,\;\cos\theta^{\rm rec}_\mu\right)$ which get contributions from each bin of separately reconstructed $\mu^-$ and $\mu^+$ events. While estimating these density distributions of ICAL sensitivity for the NUNM scenario, we do not consider the constant contribution to $\Delta\chi^2$ from the pull penalty term mentioned in eq.~(\ref{Eq:5.1}). However, we do minimize the $\Delta\chi^2$ $\left[\rm GeV^{-1}\, sr^{-1}\right]$ over systematic uncertainties. As far as standard oscillation parameters are concerned, we keep them fixed at values given in table \ref{Tab:1}.

In figure~\ref{fig:4}, we show the distributions of $\Delta\chi^2_-$ $\left[\rm GeV^{-1}\, sr^{-1}\right]$ and $\Delta\chi^2_+$ $\left[\rm GeV^{-1}\, sr^{-1}\right]$ in the plane of $\left(E^{\rm rec}_\mu,\;\cos\theta^{\rm rec}_\mu\right)$ while considering the two values NUNM parameter $\left(\alpha_{32} = \pm\, 0.1\right)$ for a demonstartion purpose. We perform this analysis using ICAL simulated data for an exposure of $500$ kt$\cdot$yr. The sensitivity for $\mu^-$ is larger than that for $\mu^+$, and this could be because of the larger statistics and significant matter effect (for NO) for the case of $\mu^-$. In all the panels, ICAL sensitivity for NUNM is higher for regions of larger baselines which correspond to the core-passing neutrinos. A significant sensitivity is obtained around the oscillation valley region, i.e., $5\;{\rm GeV}\; \le E^{\rm rec}_\mu\;\le 17\;{\rm GeV}$ and $\cos\theta^{\rm rec}_\mu < -0.4$.

%=============================%
\subsection{Impact of oscillation parameter marginalization in constraining $\alpha_{32}$}
\label{subsec:minimization}
%=============================%
Over the past few decades, the precision on the three-flavor neutrino oscillation parameters have improved significantly. However, some oscillation parameters still have large uncertainties. The next-generation neutrino oscillation experiments aim to measure the value of $\delta_{\rm CP}$, determine the octant of $\theta_{23}$, and resolve the issue of neutrino mass ordering. In ICAL, more than $98\%$ of the muon events would be contributed by $\nu_\mu$ survival channel and the contribution from the appearance channel is very minimal. Therefore, the ICAL sensitivities do not depend much on the value of $\delta_{\rm CP}$~\cite{Gandhi:2007td, Blennow:2012gj,ICAL:2015stm}. However, the performance of ICAL depends on the choice of $\sin^2\theta_{23}$, $\Delta m^2_{32}$, and the neutrino mass ordering. Therefore, in this paper, in order to estimate the possible upper bounds on the NUNM parameter $\alpha_{32}$, we minimize the $\Delta\chi^2$ in the fit over $\sin^2\theta_{23}\in\left[0.36,\,0.66\right],\; \Delta m^2_{32}\in\left[2.1,\,2.6\right]\times 10^{-3}\;{\rm eV^2}$, and both the choices of mass ordering. Here, the prospective MC data is generated using the benchmark values of the neutrino oscillation parameters as given in table~\ref{Tab:1} assuming NMO as true neutrino mass ordering. Note that the value of $\delta_{\rm CP}$ is kept fixed at zero both in theory and MC data.

%=============================%
\begin{figure}[t]
\centering
\includegraphics[width=0.75\linewidth]{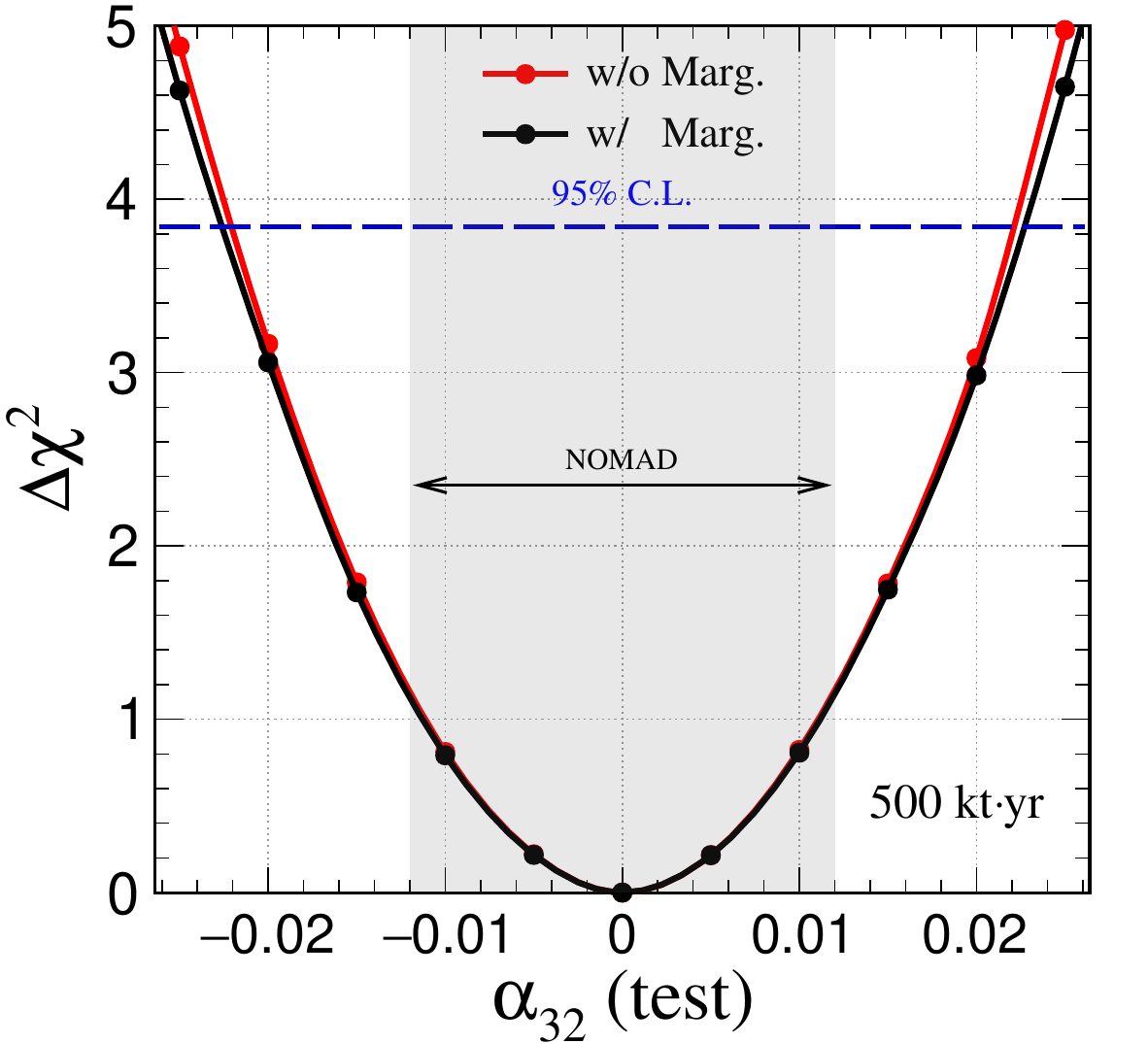}
\mycaption{The possible constraints on the NUNM parameter $\alpha_{32}$ using $500$ kt$\cdot$yr of simulated ICAL data with the CID capability. The black (red) curve shows the result with (without) minimization of the $\Delta \chi^2$ over $\sin^2\theta_{23}$, $|\Delta m^2_{32}|$, and both the choices of mass orderings in the fit. The MC data is generated for the SI case ($\alpha_{32} = 0$) using the benchmark values of oscillation parameters as given in table~\ref{Tab:1}. The gray band shows the present limit on $\alpha_{32}$ at 95\% C.L. as obtained from the SBL NOMAD experiment~\cite{NOMAD:2001xxt, NOMAD:2003mqg}.}
\label{fig:5}
\end{figure}
%=============================%

In figure~\ref{fig:5}, the black curve shows the constraints on the NUNM parameter $\alpha_{32}$ using $500$ kt$\cdot$yr of simulated ICAL data with the CID capability. As discussed above, while obtaining these sensitivities, the MC data is generated assuming $\alpha_{32} = 0$ (SI case) using the benchmark values of oscillation parameter given in table~\ref{Tab:1} and we minimize the $\Delta\chi^2$ over all the relevant oscillation parameters $\sin^2\theta_{23}$, $|\Delta m^2_{32}|$, and both the choices of mass orderings in the fit. For the first time, in a model independent fashion, we evaluate the sensitivity for the NUNM parameter $\alpha_{32}$ as: $-0.022 \le \alpha_{32} \le 0.022$  at 95\% C.L. with 1 d.o.f. assuming true NMO. The red curve in figure~\ref{fig:5} is obtained in fixed-parameter scenario where we do not marginalize over relevant oscillation parameter in the fit. We observe that the red curve almost overlaps with the black curve suggesting that the marginalization over oscillation parameter does not change our sensitivity. This implies that the projected constraints on $\alpha_{32}$ from ICAL are quite robust against the present uncertainties in the three-flavor neutrino oscillation parameters. We also estimate the future constraints on $\alpha_{32}$ as $-0.025 \le \alpha_{32} \le 0.025$  at 95\% C.L. assuming true IMO where we switch from NMO to IMO in our analysis following the prescription as described in section~\ref{sec:atm-tool}.

%=============================%
\begin{figure}[t]
\centering
\includegraphics[width=0.75\textwidth]{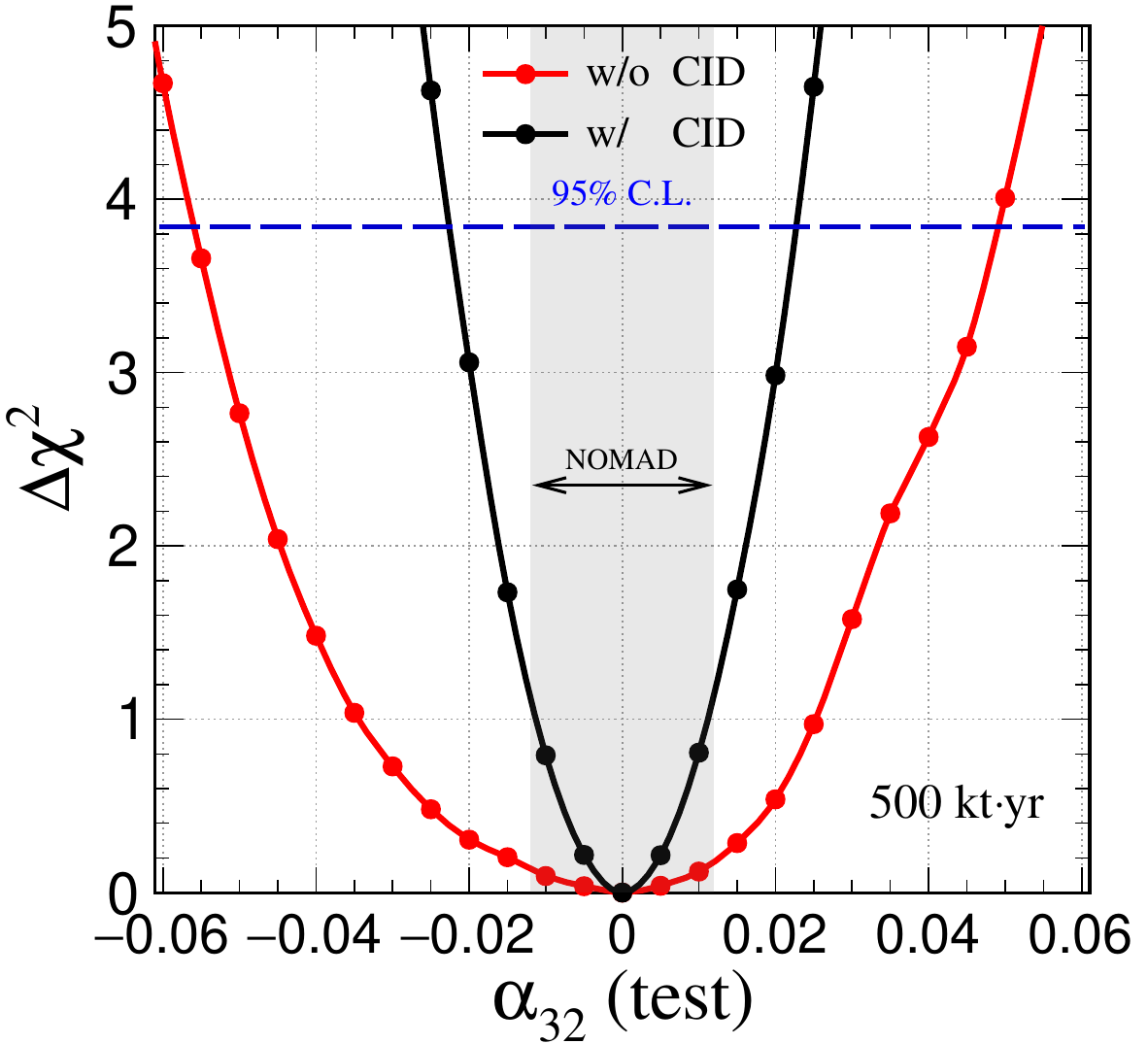}
\mycaption{The possible constraints on the NUNM parameter $\alpha_{32}$ using $500$ kt$\cdot$yr of simulated ICAL data. The black (red) curve shows the result with (without) the CID capability of ICAL. In the fit, we minimize the $\Delta \chi^2$ over $\sin^2\theta_{23}$, $|\Delta m^2_{32}|$, and both the choices of mass orderings. The MC data is generated for the SI case ($\alpha_{32} = 0$) using the benchmark values of oscillation parameters as given in table~\ref{Tab:1}. The gray band shows the present limit on $\alpha_{32}$ at 95\% C.L. as obtained from the SBL NOMAD experiment~\cite{NOMAD:2001xxt, NOMAD:2003mqg}.}
\label{fig:6}
\end{figure}
%=============================%

%=============================%
\subsection{Advantage of having charge identification (CID) capability}
%=============================%
The magnetic field of 1.5 T would enable ICAL to identify $\mu^-$ and $\mu^+$ events separately by measuring the opposite directions of curvatures of their tracks. This CID capability of ICAL would help in distinguishing parent neutrinos ($\nu_\mu$) and antineutrinos ($\bar\nu_\mu$) which experience Earth's matter effects in a different fashion for a given mass ordering. The CID capability helps in preserving the different Earth's matter effects in neutrino and antineutrino modes separately, which in turn enhances the sensitivity of ICAL in measuring the standard three-flavor oscillation parameters and also in probing the various new-physics scenarios driven by Earth's matter effect.

In figure~\ref{fig:6}, we show the advantage of having the CID capability of ICAL in constraining the NUNM parameter $\alpha_{32}$. In this figure, we plot the $\Delta\chi^2$ (see eq.~(\ref{Eq:5.4})) as a function of $\alpha_{32}$ in the fit where the true value $\alpha_{32}$ is zero in the MC data. Here, we minimize the $\Delta\chi^2$ over $\sin^2\theta_{23}$, $|\Delta m^2_{32}|$, and both the mass orderings in the fit assuming true NMO. The black curve represents the sensitivity of ICAL with the CID capability, whereas the red curve corresponds to the sensitivity without CID. We observe that in the absence of CID, the constraint on $\alpha_{32}$ deteriorates to $-0.056 \le \alpha_{32} \le 0.049$ at $95\%$ C.L. with 500 kt$\cdot$yr exposure as compared to the constraint $-0.022 \le \alpha_{32} \le 0.022$ that we have in the presence of CID. This happens because in the absence of CID, the reconstructed $\mu^-$ and $\mu^+$ events get added up in each bin diluting their matter effect information which in turn deteriorates the sensitivity towards $\alpha_{32}$. Also note that in the absence of CID, the $\alpha_{32}$ sensitivity becomes asymmetric for positive and negative values of $\alpha_{32}$ (test). So overall, the numbers in figure~\ref{fig:6} indicate an improvement of around 60\% in the sensitivity towards $\alpha_{32}$ in the presence of CID.

%=============================%
\subsection{Constraints on $\alpha_{32}$ as a function of true $\sin^2\theta_{23}$}
%=============================%
%=============================%
\begin{figure}[t]
\centering
\includegraphics[width=0.75\linewidth]{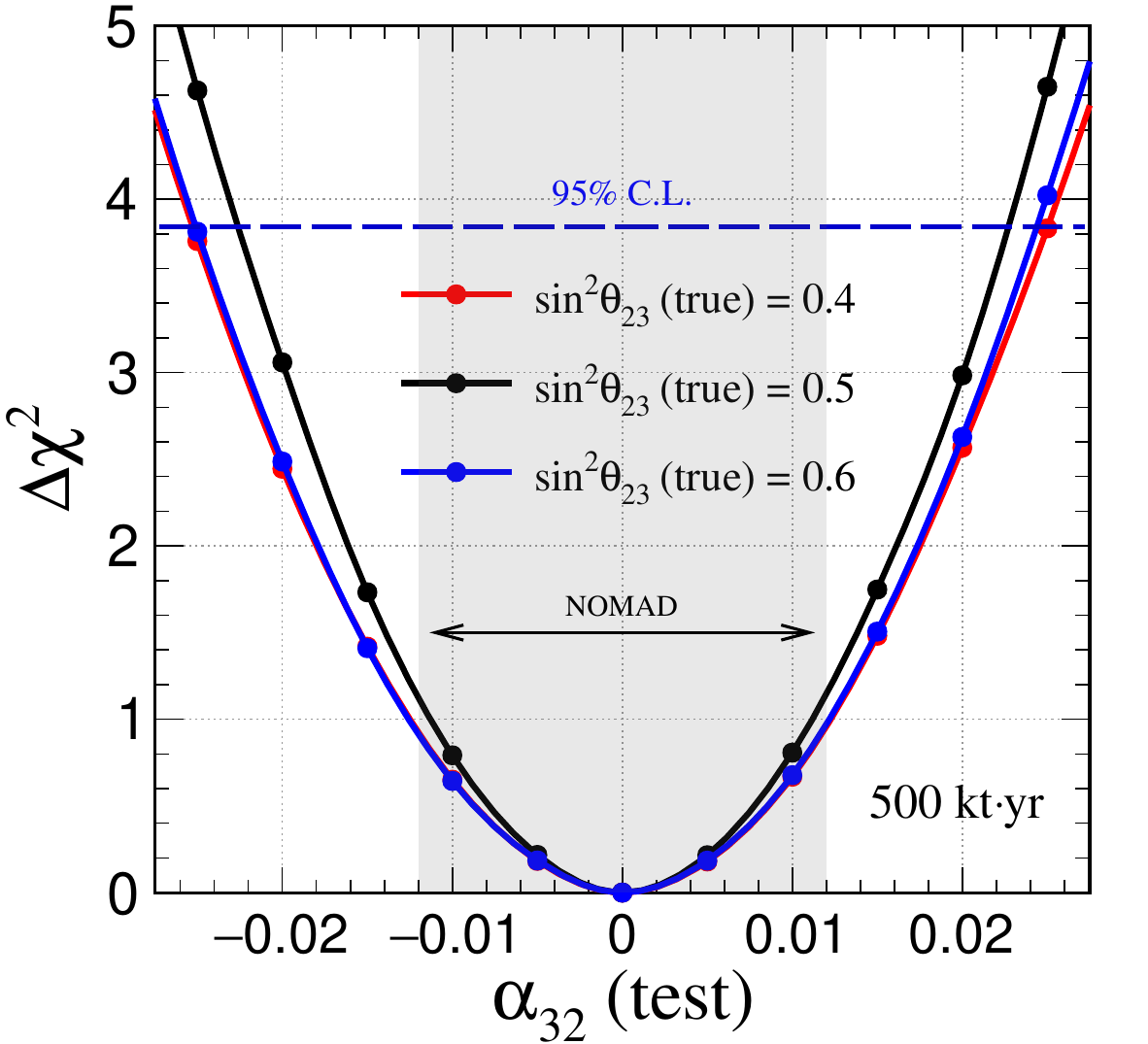}
\mycaption{The possible constraints on the NUNM parameter $\alpha_{32}$ using $500$ kt$\cdot$yr of simulated ICAL data with the CID capability, while considering three different true values of $\sin^2\theta_{23}$: $0.4$ (red curve), $0.5$ (black curve), and $0.6$ (blue curve). In the fit, we minimize the $\Delta \chi^2$ over $\sin^2\theta_{23}$, $|\Delta m^2_{32}|$, and both the choices of mass orderings. The MC data is generated for the SI case ($\alpha_{32} = 0$) using the benchmark values of oscillation parameters as given in table~\ref{Tab:1}. The gray band shows the present limit on $\alpha_{32}$ at 95\% C.L. as obtained from the SBL NOMAD experiment~\cite{NOMAD:2001xxt, NOMAD:2003mqg}.}
\label{fig:7}
\end{figure}
In figure~\ref{fig:7}, we examine the impact of uncertainty of the true value of $\theta_{23}$ on ICAL sensitivity to constrain $\alpha_{32}$ with $500$ kt$\cdot$yr exposure. The black curve corresponds to the sensitivity using MC data with maximal mixing, i.e., $\sin^2\theta_{23}$ (true) $ = 0.5$, whereas the red and blue curves represent that with non-maximal mixings, i.e., $\sin^2\theta_{23}$ (true) $ = 0.4$ and $0.6$, respectively. Here, we minimize $\Delta\chi^2$ over $\sin^2\theta_{23}$, $|\Delta m^2_{32}|$, and mass ordering while considering NMO as the true scenario. The ICAL sensitivity for $\alpha_{32}$ gets deteriorated by $\approx 14\%$ when we consider a non-maximal mixing of $\theta_{23}$ (red and blue curves) in MC data as compared to that for maximal mixing (black curve). The sensitivity for lower ($\sin^2\theta_{23} = 0.4$) and higher ($\sin^2\theta_{23} = 0.6$) octant is also similar. This happens because $P\left(\nu_\mu \to \nu_\mu\right)$ depends upon $\sin^2{2\theta_{23}}$ at the leading order, enhancing the flavor transition probability, which is described in more detail in appendix~\ref{Eq:A1}.
%=============================%

%=============================%
\subsection{One-to-one comparison of ICAL with the future long-baseline setups}
\label{sec:Comp.}
%=============================%
The upcoming high-precision LBL experiments DUNE~\cite{DUNE:2015lol,DUNE:2020lwj,DUNE:2020ypp,DUNE:2020jqi,DUNE:2020fgq,DUNE:2021mtg,DUNE:2021cuw} and T2HK~\cite{Abe:2016srs,Abe:2018ofw} are expected to provide stringent constraints on the NUNM parameters~\cite{Blennow:2016jkn,Escrihuela:2016ube,Agarwalla:2021owd}. In this section, we provide a comparison between the future sensitivities obtained from these LBL experiments and the INO-ICAL atmospheric neutrino experiment. For this purpose, we perform a detailed simulation study to estimate the limits on the NUNM parameter $\alpha_{32}$ using the LBL experiments DUNE~\cite{DUNE:2021cuw} and T2HK~\cite{Abe:2016srs} in isolation and combination. To simulate the performance of DUNE having a baseline of 1300 km, we consider a total exposure of 480 kt$\cdot$MW$\cdot$yr which corresponds to a 1.2 MW beam of protons of 120 GeV and a 40 kt liquid-argon time-projection chamber (LArTPC) as a far detector collecting data for 10 years equally divided in neutrino (5 years) and antineutrino (5 years) modes. To simulate the prospective data of T2HK with a baseline of 295 km, we consider a total exposure of 2431 kt$\cdot$MW$\cdot$yr which is obtained with a 1.3 MW beam of protons of 30 GeV and a 187 kt water Cherenkov far detector collecting data for 10 years with 2.5 years of neutrino run and 7.5 years of antineutrino run. We perform the necessary simulation for these two LBL experiments using the publicly available GLoBES software~\cite{Huber:2004ka, Huber:2007ji}, extended with the MonteCUBES package~\cite{Blennow:2009pk}. For both these experiments, we use the benchmark values of the oscillation parameters as given in table~\ref{Tab:1} assuming true NMO to generate the MC data with the NUNM parameter $\alpha_{32} = 0$, and consider the non-zero positive and negative real values of  $\alpha_{32}$ in the fit. For both these LBL setups, we minimize the $\Delta \chi^2$ over $\sin^2\theta_{23}$, $|\Delta m^2_{32}|$, and both the choices of mass orderings in the fit. Similar to the ICAL analysis, we consider $\delta_{\rm CP} = 0$ both in data and fit while obtaining the sensitivities for these LBL setups.
 
%=============================%
\begin{figure}[ht]
\centering
\includegraphics[width=0.75\linewidth]{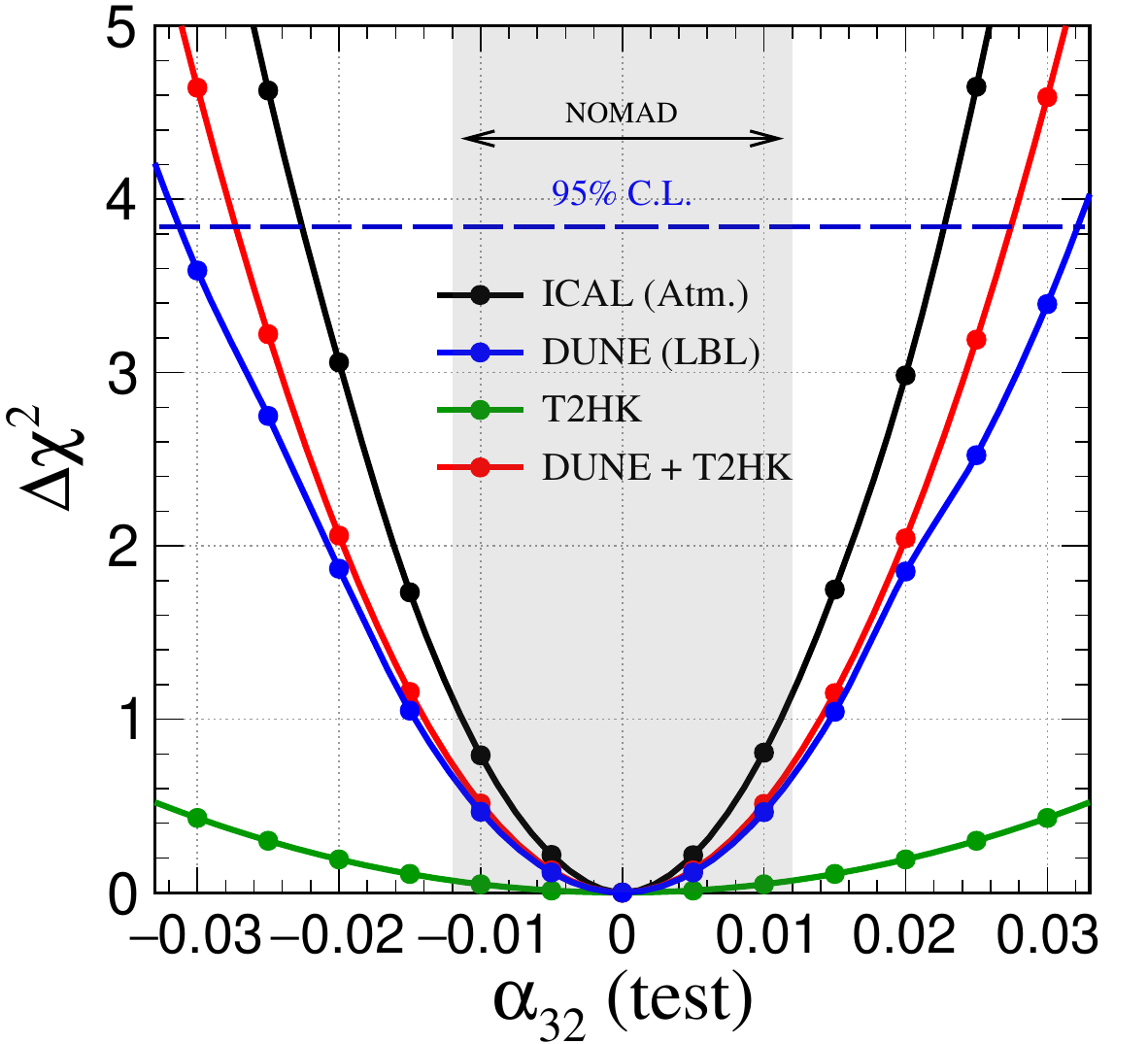}
\mycaption{The possible constraints on the NUNM parameter $\alpha_{32}$ using $500$ kt$\cdot$yr of simulated atmospheric data at ICAL with the CID capability (see black curve). The blue (green) curve shows the $\alpha_{32}$ sensitivity of DUNE (T2HK) long-baseline setup assuming a total exposure of $480$ kt$\cdot$MW$\cdot$yr ($2431$ kt$\cdot$MW$\cdot$yr) in 10 years. For DUNE, we assume 5 years of neutrino run and 5 years of antineutrino run. For T2HK, we consider 2.5 years of neutrino run and 7.5 years of antineutrino run. The red curve portrays the $\alpha_{32}$ sensitivity using the combined analysis of the long-baseline MC data from DUNE and T2HK. For all these experimental setups, we minimize the $\Delta \chi^2$ over $\sin^2\theta_{23}$, $|\Delta m^2_{32}|$, and both the choices of mass orderings. The MC data is generated for the SI case ($\alpha_{32} = 0$) using the benchmark values of oscillation parameters as given in table~\ref{Tab:1} and assuming NMO as true mass ordering. We consider $\delta_{\rm CP} = 0$ both in data and fit while obtaining the results for all these setups. The gray band shows the present limit on $\alpha_{32}$ at 95\% C.L. as obtained from the SBL NOMAD experiment~\cite{NOMAD:2001xxt, NOMAD:2003mqg}.}
  \label{fig:8}
\end{figure}
%=============================%

In figure~\ref{fig:8}, we show the values of minimized $\Delta\chi^2$ as a function of $\alpha_{32}$ (test) in fit. The black curve shows the $\alpha_{32}$ sensitivity of INO-ICAL atmospheric neutrino experiment considering its CID capability. The blue (green) curve reveals the $\alpha_{32}$ sensitivity for the standalone DUNE (T2HK) LBL setup. The red curve depicts the sensitivity for the combined analysis of DUNE and T2HK. We observe that the sensitivity from the combined DUNE + T2HK analysis is better than their individual performance. Owing to the smaller baseline, and hence, less matter effect, the sensitivity of T2HK (see green curve) is significantly less as compared to DUNE (see blue curve). We also notice in this one-to-one comparison that the sensitivity of ICAL (see black curve) is slightly better than DUNE and DUNE + T2HK owing to the large matter effects via NC interactions experienced by atmospheric neutrinos at ICAL. We tabulate the constraints on the NUNM parameter $\alpha_{32}$ obtained from these various experiments at 95\% C.L. in table~\ref{Tab:3}. 

%=============================%
\begin{table}[t]
  \centering
  \begin{tabular}{|c|c|c|c|}
    \hline
    ICAL & DUNE & T2HK & DUNE + T2HK\\
    \hline
    [-0.022, 0.022] & [-0.031, 0.032] & [-0.089, 0.089] & [-0.027, 0.027]\\
    \hline 
  \end{tabular}
  \mycaption{A comparison between the expected constraints on the NUNM parameter $\alpha_{32}$ at $95$\% C.L. obtained from $500$ kt$\cdot$yr atmospheric neutrino exposure of ICAL (first column), 480 kt$\cdot$MW$\cdot$yr exposure of DUNE LBL setup (second column), and 2431 kt$\cdot$MW$\cdot$yr exposure of T2HK LBL setup (third column). The last column shows the sensitivity of $\alpha_{32}$ obtained from the combined analysis of DUNE and T2HK LBL data. For all the experimental setups, we minimize the $\Delta \chi^2$ over $\sin^2\theta_{23}$, $|\Delta m^2_{32}|$, and both the choices of mass orderings in the fit. The prospective MC data is generated using the benchmark values of oscillation parameters given in table~\ref{Tab:1}. We consider $\delta_{\rm CP} = 0$ both in data and fit.}
\label{Tab:3}
\end{table}
%=============================%

So far, in this paper, we treat $\alpha_{32}$ as a real parameter with positive and negative values. In other words, we consider the NUNM phase $\phi_{32}$ associated with $\alpha_{32}$ (in the convention of $\alpha_{32}\equiv |\alpha_{32}|e^{-i\phi_{32}}$) to be 0 and $\pi$. Now, we study the impact of the NUNM phase $\phi_{32}$ on the future sensitivities towards $|\alpha_{32}|$ which can be obtained from the above-mentioned LBL and atmospheric neutrino experiments. To see the impact of the NUNM phase $\phi_{32}$ on the INO-ICAL analysis, we repeat our simulations for INO-ICAL by performing the minimization of the $\Delta \chi^2$ over all possible values of $\phi_{32} \in \left[-\pi,\pi\right]$ in the fit while keeping all the standard three-flavor neutrino oscillation parameters fixed in the fit for computational ease. We obtain a new constraint of $|\alpha_{32}| \le 0.12$ at 95\% C.L. using 500 kt$\cdot$yr exposure of INO-ICAL. We also repeat the simulations for DUNE and T2HK in isolation and combination by performing the minimization of the $\Delta \chi^2$ over $\phi_{32}$ in the fit while keeping the standard three-flavor neutrino oscillation parameters fixed in the fit. While studying the impact of $\phi_{32}$ on the LBL setups, we also consider $\delta_{\rm CP} = 0$ both in data and fit as we assume for ICAL. In table~\ref{Tab:5}, we compare the sensitivities for $|\alpha_{32}|$ obtained from DUNE, T2HK, and DUNE + T2HK with that from ICAL by performing the minimization over $\phi_{32}$ in the fit. We observe that owing to Earth's matter effects, INO-ICAL has slightly better sensitivity for $|\alpha_{32}|$ as compared to DUNE, T2HK, and their combination even after the marginalization over the phase $\phi_{32}$. 
%=============================%
\begin{table}[t]
  \centering
  \begin{tabular}{|c|c|c|c|}
    \hline
    ICAL & DUNE & T2HK & DUNE + T2HK\\
    \hline
    $<$ 0.12 & $<$ 0.21 & $<$ 1.08 & $<$ 0.20\\
    \hline 
  \end{tabular}
  \mycaption{A comparison between the expected constraints on $|\alpha_{32}|$ at $95$\% C.L. obtained after marginalizing over the NUNM phase  $\phi_{32} \in \left[-\pi,\pi\right]$ using $500$ kt$\cdot$yr exposure of INO-ICAL atmospheric neutrino experiment (first column), 480 kt$\cdot$MW$\cdot$yr exposure of DUNE LBL setup (second column), and 2431 kt$\cdot$MW$\cdot$yr exposure of T2HK LBL setup (third column). The last column shows the sensitivity for $|\alpha_{32}|$ obtained from the combined analysis of DUNE and T2HK. For all the experimental setups, we keep the value of the oscillation parameters fixed  both in data and fit at their benchmark values as given in table~\ref{Tab:1}. We also consider $\delta_{\rm CP} = 0$ both in data and fit.}
\label{Tab:5}
\end{table}
%=============================%
\subsection{Sensitivities in $|\alpha_{32}|$ and $\phi_{32}$ plane}
\label{sec:Limits_DCP}
\begin{figure}[t]
\centering
\includegraphics[width=0.48\textwidth]{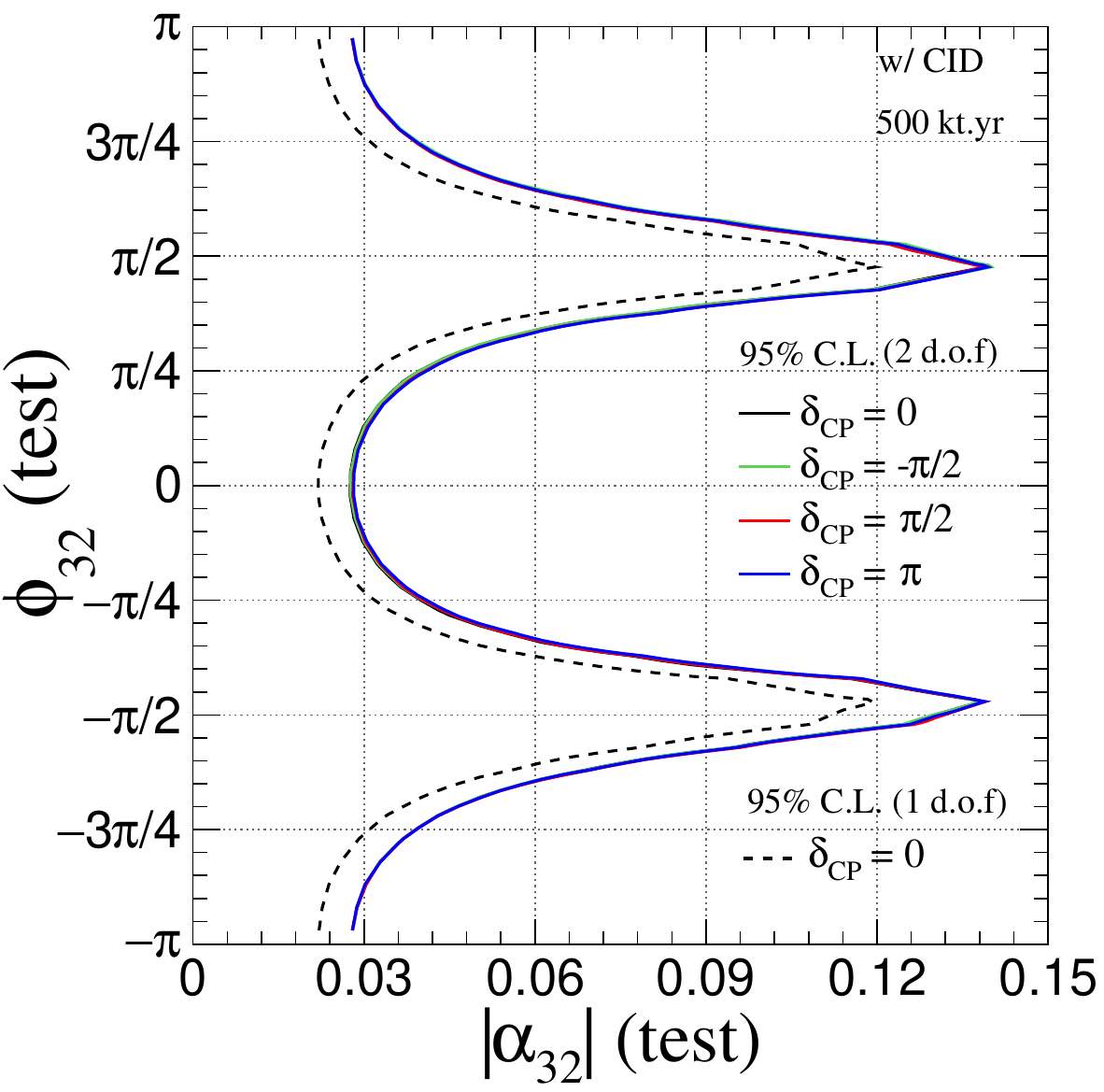}
\quad
\includegraphics[width=0.48\textwidth]{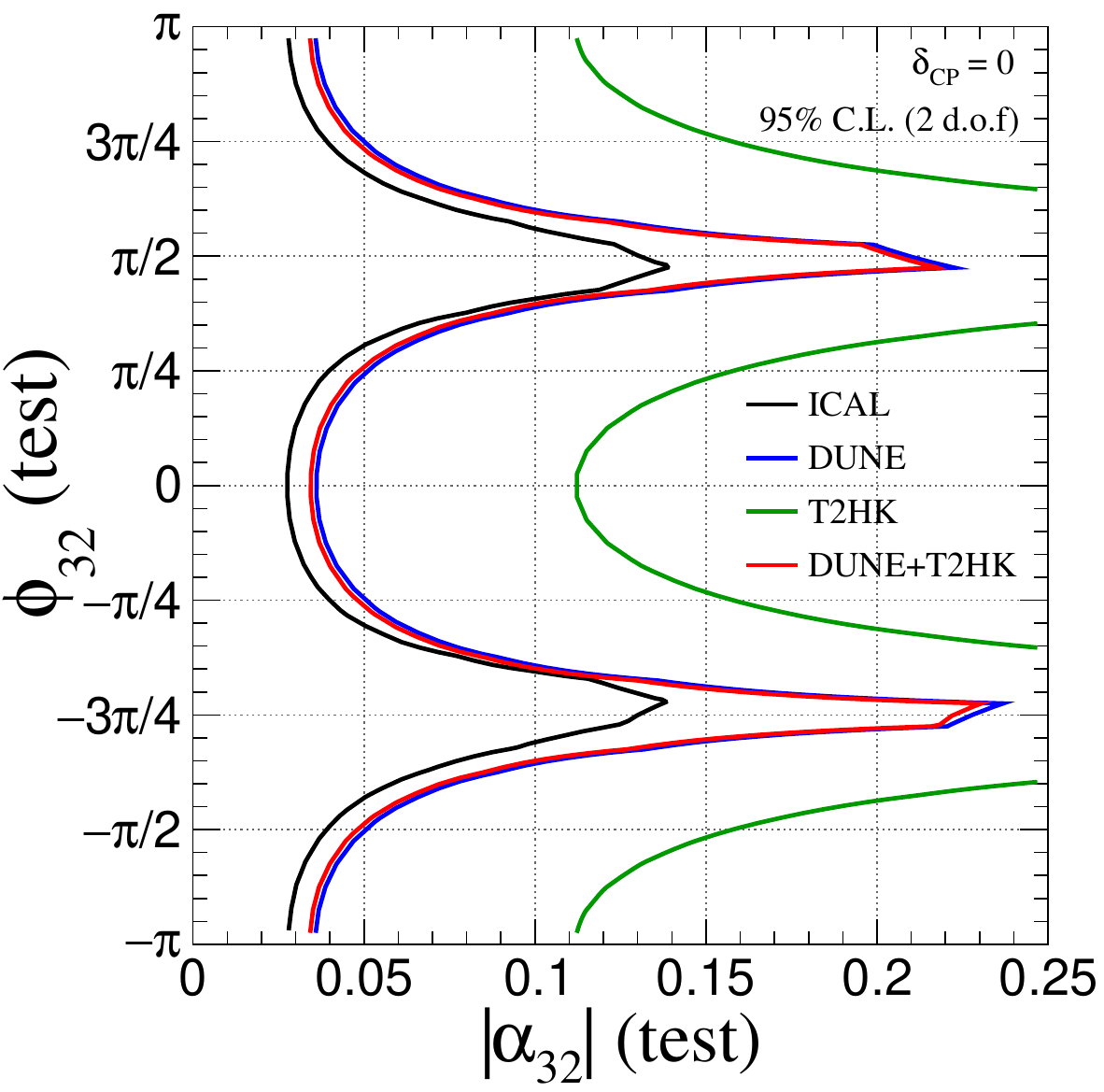}
\mycaption{The left panel shows the expected constraints on the NUNM parameter $\alpha_{32}$ on the plane of $\left(|\alpha_{32}|,\, \phi_{32}\right)$ using $500$ kt$\cdot$yr exposure of the simulated ICAL data with the CID capability. We consider four different true values of $\delta_{\rm CP}$: $0$ (black curve), $-\pi/2$ (green curve), $\pi/2$ (red curve) and $\pi$ (blue curve) at $95\%$ C.L. (2 d.o.f), while keeping $\delta_{\rm CP}$ fixed in the fit at its respective true values. In the fit, we also keep all other oscillation parameters fixed at their true values. Additionally, the black dashed curve shows the limits on $|\alpha_{32}|$ for 1 d.o.f. at $95\%$ confidence level. The right panel shows the expected sensitivities for future long baseline experiments in the plane of $|\alpha_{32}|$ and $\phi_{32}$. The blue (green) curve shows the sensitivity for DUNE (T2HK) long-baseline setup assuming a total exposure of $480$ kt$\cdot$MW$\cdot$yr ($2431$ kt$\cdot$MW$\cdot$yr) in 10 years. For DUNE, we assume 5 years of neutrino run and 5 years of antineutrino run. For T2HK, we consider 2.5 years of neutrino run and 7.5 years of antineutrino run. The red curve portrays the sensitivity using the combined analysis from DUNE and T2HK. We consider $\delta_{\rm CP} = 0$ both in MC data and fit for all these experimental setups. For both the panels, the MC data is generated for the SI scenario ($|\alpha_{32}| = 0$, $\phi_{32}=0$) using the benchmark values of oscillation parameters as given in table~\ref{Tab:1} and assuming NMO as true mass ordering.}
\label{fig:NUNMdcp}
\end{figure}

In table \ref{Tab:5}, we show that how the sensitivity of standalone ICAL for constraining $|\alpha_{32}|$ gets impacted by minimizing the $\Delta\chi^2$ over the NUNM phase $\phi_{32}\in \left[-\pi,\,\pi\right]$. In figure~\ref{fig:NUNMdcp}, we show the sensitivities of ICAL in the $|\alpha_{32}|$ and $\phi_{32}$ plane. The solid black curves indicate that the sensitivity for $|\alpha_{32}|$ gets significantly deteriorated for $\phi_{32}\simeq\pm\,\pi/2$ in the fit. This can be explained from the approximate expression of the $\nu_\mu\to\nu_\mu$ survival probability derived in appendix~\ref{app:A3}, where we observe that the leading term containing $|\alpha_{32}|$ has dependence of $\cos\phi_{32}$. Therefore, for $\phi_{32}\simeq \pm\,\pi/2$, the contribution from $|\alpha_{32}|$ in the survival probability almost vanishes, reducing its sensitivity near those values.

%=============================%
%=============================%
In the left panel, we show the sensitivity curves for four different true values of $\delta_{\rm CP}$, namely, $0$ (black curve), $-\pi/2$ (green curve), $\pi/2$ (red curve) and $\pi$ (blue curve) at $95$\% C.L. (2 d.o.f), while $\delta_{\rm CP}$ in fit is also kept fixed at its respective true value. We observe that our result is robust under all possible values of $\delta_{\rm CP}$. Therefore, in future, if the $\delta_{\rm CP}$ found to be non-zero with a precise value, it would not impact the constraint placed by ICAL. This happens because survival channel of muon neutrino $P\left(\nu_\mu \to \nu_\mu\right)$ and $P\left(\bar\nu_\mu \to \bar\nu_\mu\right)$ contributes about $98\%$ of the reconstructed $\mu^-$, and $\mu^+$ events at the ICAL detector, respectively. In the approximate expression of $\nu_\mu\to\nu_\mu$ survival probability, as shown in ref.~\cite{Akhmedov:2004ny}, $\delta_{\rm CP}$ appears in terms of $\cos\delta_{\rm CP}$ which is suppressed by a factor of $\frac{\Delta m^2_{21}}{\Delta m^2_{31}}\sin\theta_{13}\simeq 0.005$. Therefore, the effect of $\delta_{\rm CP}$ is negligible irrespective of any new physics scenario. Note that we perform this analysis in a fixed oscillation parameters scenario, as we find in figure~\ref{fig:5} that there is no significant impact of minimization of $\Delta\chi^2$ over $\sin^2\theta_{23}$, $|\Delta m^2_{32}|$ and both choices of mass orderings in the fit. We also show the sensitivity for 1 d.o.f at $95\%$ C.L.  as indicated by the black dashed curve, in order to have a comparison with $|\alpha_{32}|$ sensitivity shown in table~\ref{Tab:5} for 1D analysis discussed in the previous section.

In the right panel, we compare the sensitivity of ICAL with that of the next-generation long-baseline experiments DUNE, T2HK, and their combination in the $|\alpha_{32}|$ and $\phi_{32}$ plane at 95\% C.L. (2 d.o.f.). Here, we consider the true value of $\delta_{\rm CP} = 0$ for all the experimental setups. Similar to the observation made in section~\ref{sec:Comp.}, we found that the ICAL sensitivity (shown by the black curve) is slightly better than DUNE and T2HK setups, individually and in combination.

%=============================%
\input{plots/Limits/Limits.tex}
%=============================%

%=============================%
\section{Concluding remarks}
\label{sec:CR}
%=============================%
The experimental evidence of the mass-induced neutrino flavor transitions places a dent on the Standard Model (SM) of particle physics and this basically opens up a gateway to the phenomenology of beyond the SM (BSM). The mixing among three light active neutrinos is described by unitary matrix. Since over the past few decades, the precision of neutrino oscillation parameters have improved significantly, it is natural to attempt to test the unitarity of mixing matrix. In this article, we study the non-unitary neutrino mixing (NUNM) scenario using atmospheric neutrinos over a wide range of baselines in a multi-GeV range of energies at the upcoming INO-ICAL detector. Here, our cornerstone is exploring the non-unitary neutrino mixing parameter $\alpha_{32}$ through mass-induced neutrino oscillations. We consider only the real values of $\alpha_{32}$ and show that ICAL can place a stringent limit on it. Here, we study the possibility of non-unitary neutrino mixing among the three SM neutrinos. We discuss in detail how the NUNM scenario can uphold the potential arising due to neutral-current interactions and can make a significant alteration in the neutrino oscillation probabilities. We explore the impact of the NUNM parameter $\alpha_{32}$ on the $\nu_\mu \to \nu_\mu$ survival channel and derive an approximate analytical expression for an effective two-neutrino mixing in 2-3 sector (see appendix~\ref{app:A2}). We demonstrate such an effect using  oscillogram of $\nu_\mu$ survival probability.

We simulate the effect of the NUNM parameter $\alpha_{32}$ on the distribution of reconstructed muon events for the upcoming 50 kt ICAL with an exposure of 10 years. Here, we perform a $\Delta\chi^2$ analysis to estimate the new physics sensitivity of the INO-ICAL detector. In this work, we evaluate the sensitivity to place a limit on the NUNM parameter $\alpha_{32}$ using an atmospheric neutrino experiment like ICAL. We discuss the advantages of using the charge-identification feature at ICAL both at the event level and $\Delta\chi^2$ analysis. We find that the $\alpha_{32}$ sensitivity is robust against the minimization of $\Delta\chi^2$ over the uncertainties of atmospheric oscillation parameters. We further explore the impact of octant uncertainty of $\theta_{23}$ (true) while constraining the NUNM parameter $\alpha_{32}$. Since there is literature that discusses the NUNM scenario for the mass-induced neutrino oscillations using the long-baseline experimental data and the proposed long-baseline simulated data, we perform a one-to-one comparison of the sensitivities of the future long-baseline experiments such as DUNE, T2HK as well as their combination (DUNE + T2HK) with that of ICAL. 
In figure~\ref{fig:nunma32limits}, we show a comparison of the expected bounds on $|\alpha_{32}|$ using ICAL with its CID capability and the limits from the current and future experiments. Figure~\ref{fig:nunma32limits} reveals that if the NUNM parameter $\alpha_{32}$ is assumed to be real, then the proposed ICAL detector would be able to provide a comparable and complementary constraint using atmospheric neutrinos as compared to the current limits obtained from the SBL NOMAD experiment and electroweak precision data. We also observe that the complex phase $\phi_{32}$ associated with $\alpha_{32}$ affects the ICAL sensitivity and makes the limit weaker by almost an order of magnitude.

For probing the BSM physics scenarios, it is essential to make precise measurements of oscillation parameters, and such a process will be enhanced with the detection of $\nu_\tau$ events. We believe that the next-generation neutrino detectors will be capable of observing the charged-current tau events via the tau neutrino appearance channel. Such an event analysis will undoubtedly enhance the robustness of three-neutrino unitary mixing. In fact, a thorough analysis of currently acquired high-precision atmospheric neutrino data at Super-K, IceCube, DeepCore, and ORCA can definitely shed a light on the test for unitary neutrino mixing.

%=============================%
\section*{Acknowledgements}
\label {subsec:ackw}
%=============================%
We thank the members of the INO-ICAL collaboration, and the anonymous referee for their valuable comments and constructive inputs. We sincerely thank A. Dighe, S. Goswami, S. Choubey, and P. Swain, A. Giarnetti for their useful comments and suggestions.  We acknowledge the support of the Department of Atomic Energy (DAE), Govt. of India, under Project Identification No. RTI4002. S.K.A. gets supported by the DST/INSPIRE Research Grant [IFA-PH-12] from the Department of Science and Technology (DST), Govt. of India, and the Young Scientist Project [INSA/SP/YSP/144/2017/1578] from the Indian National Science Academy (INSA). S.K.A. acknowledges the financial support from the Swarnajayanti Fellowship Research Grant (No. DST/SJF/PSA-05/2019-20) provided by the Department of Science and Technology (DST), Govt. of India, and the Research Grant (File no. SB/SJF/2020-21/21) provided by the Science and Engineering Research Board (SERB) under the Swarnajayanti Fellowship by the DST, Govt. of India. S. Sahoo thanks the organizers of the XXV DAE-BRNS symposium 2022, IISER, Mohali, India, from 12th to 16th December 2023, for providing him an opportunity to give a talk to present the preliminary results from this work. We acknowledge the Sim01: High-Performance Computing facilities at Tata Institute of Fundamental Research, Mumbai for performing the numerical simulations.

%=============================%
\begin{appendix}
%=============================%
\section{A brief discussion on lower-triangular formulation of NUNM }
\label{app:A1}
\noindent
Let's consider a scenario where `$n$' numbers of neutrinos are mixing with each other via a global mixing matrix $\widetilde{U}$ that is an $n\times n$ square matrix. Recalling eq.~(\ref{Eq:2.3}) to represent the mixing matrix as follows:
\begin{align}
  \widetilde{U}^{n\times n} & = \left(
  \begin{array}{cc}
    N & S \\
    V & T
  \end{array}
  \right) \equiv \left(
  \begin{array}{c|c}
    N^{3 \times 3} & S^{3 \times (n-3)} \\
    \hline
    V^{(n-3) \times 3} & T^{(n-3) \times (n-3)}
  \end{array}
  \right)\,.
  \label{Eq:A1.1}
\end{align}

\noindent
Here, $N$, $S$, $V$ and $T$ are the elements of $\widetilde{U}^{n \times n}$ matrix in the form of block matrices. Now, transforming neutrino mass basis to flavor basis, 

\begin{align}
  \begin{bmatrix}
    \nu_e    \\
    \nu_\mu  \\
    \nu_\tau \\ \hline
    . \\
    . \\
    . \\
    . 
  \end{bmatrix}_{n \times 1} & = \hspace{0.5cm}
  \left[\begin{array}{ccc|cccc}
    \widetilde{U}_{e1}    & \widetilde{U}_{e2}    & \widetilde{U}_{e3}    & \ldots & \ldots & \ldots & \ldots\\
    \widetilde{U}_{\mu1}  & \widetilde{U}_{\mu2}  & \widetilde{U}_{\mu3}  & \ldots & \ldots & \ldots & \ldots\\
    \widetilde{U}_{\tau1} & \widetilde{U}_{\tau2} & \widetilde{U}_{\tau3} & \ldots & \ldots & \ldots & \ldots\\ \hline
    \ldots & \ldots & \ldots & \ldots & \ldots & \ldots & \ldots \\
    \ldots & \ldots & \ldots & \ldots & \ldots & \ldots & \ldots \\
    \ldots & \ldots & \ldots & \ldots & \ldots & \ldots & \ldots \\
    \ldots & \ldots & \ldots & \ldots & \ldots & \ldots & \ldots
  \end{array}\right]_{n \times n}\cdot\hspace{0.5cm}
  \begin{bmatrix}
    \nu_1 \\
    \nu_2 \\
    \nu_3 \\ \hline
    . \\
    . \\
    . \\
    .     
  \end{bmatrix}_{n \times 1}\,.
  \label{Eq:A1.2}
\end{align}
In this representation, there are blocks showing the current accessibility regime, i.e., we can detect $\nu_e$, $\nu_\mu$, and $\nu_\tau$ via inverse-beta decay processes, and we have precisely measured the magnitudes of solar and atmospheric mass-squared splittings ($\Delta m^2_{21}$ and $\Delta m^2_{31}$) which are related to $\nu_1$, $\nu_2$ and $\nu_3$ neutrino mass states. In the accessible neutrino energy, the Hamiltonian in mass-basis can be represented in a block matrix as follows:

\begin{align}
  H^m_{n \times n} = \frac{1}{2E} \cdot \left[\begin{array}{cc}
    \delta m^2 & 0 \\
    0 & \Delta M^2
  \end{array}
  \right]_{n \times n}
  {\rm where\;\;}\delta m^2 = \left[\begin{array}{ccc}
    0 & 0 & 0 \\
    0 & \Delta m^2_{21} & 0 \\
    0 & 0 & \Delta m^2_{31}
  \end{array}\right]_{3 \times 3},\nonumber\\
  \nonumber\\
   {\rm and\;}
  \Delta M^2= \left[\begin{array}{ccccc}
    \Delta m^2_{41} & 0 & 0 & 0 & 0\\
    0 & \Delta m^2_{51} & 0 & 0 & 0\\
    0 & 0 & \cdots & 0  & 0\\
    0 & 0 & 0 & \cdots  & 0\\
    0 & 0 & 0 & 0 & \Delta m^2_{n1}
  \end{array}\right]_{(n-3) \times (n-3)}\,.  \label{Eq:A1.3}
\end{align}  

\noindent
Here, we assume the magnitude of neutrino masses and their mass-squared splittings are very small compared to their energy ($E$), i.e., $m_i \ll E$ and $\Delta m^2_{ij} \ll E$, where $i$ and $j$ are the mass-indices.\\

\noindent
\textbf{An illustration of 3+1 neutrino mixing scenario:} By considering $n = 4$, we can introduce a new neutrino, namely sterile neutrino, with mass basis $\nu_4$ and flavor basis $\nu_s$. Using eqs.~(\ref{Eq:A1.1}) \& (\ref{Eq:A1.3}), the transition amplitudes in vacuum can be expressed as: 

\begin{align}
  \left(\begin{array}{ccc|c}
    \langle\nu_e|\nu_e\rangle & \langle\nu_e|\nu_\mu\rangle & \langle\nu_e|\nu_\tau\rangle & \langle\nu_e|\nu_s\rangle \\
    \langle\nu_\mu|\nu_e\rangle & \langle\nu_\mu|\nu_\mu\rangle & \langle\nu_\mu|\nu_\tau\rangle & \langle\nu_\mu|\nu_s\rangle \\
    \langle\nu_\tau|\nu_e\rangle & \langle\nu_\tau|\nu_\mu\rangle & \langle\nu_\tau|\nu_\tau\rangle & \langle\nu_\tau|\nu_s\rangle \\ \hline
    \langle\nu_s|\nu_e\rangle & \langle\nu_s|\nu_\mu\rangle & \langle\nu_s|\nu_\tau\rangle & \langle\nu_s|\nu_s\rangle \\
  \end{array}\right) = 
  \left[\begin{array}{cccc}
    N & & S \\
    & &   \\
    V & & T
  \end{array}\right] 
  \left[\begin{array}{cccc}
    e^{-i\frac{\delta m^2}{2E}L} & & 0 \\
    & &   \\
    0 & & e^{-i\frac{\Delta m^2_{41}}{2E}L}
  \end{array}\right] 
  \left[\begin{array}{cccc}
    N^\dagger & & V^\dagger \\
    & &   \\
    S^\dagger & & T^\dagger
  \end{array}\right] \nonumber & \\
  \nonumber \\
  =
  \left[\begin{array}{ccc|ccc}
    N\,e^{-i\frac{\delta m^2}{2E}L}\,N^\dagger \; +  S\,e^{-i\frac{\Delta m^2_{41}}{2E}L}\, S^\dagger & & & &N\,e^{-i\frac{\delta m^2}{2E}L}\,V^\dagger \; +  S\,e^{-i\frac{\Delta m^2_{41}}{2E}L}\, T^\dagger \\
    & & & &\\
    (3 \times 3) & & & & (3 \times 1) \\
    & & & &\\
    \hline
    & & & &\\
    V\,e^{-i\frac{\delta m^2}{2E}L}\,N^\dagger \; +  T\,e^{-i\frac{\Delta m^2_{41}}{2E}L}\, S^\dagger & & & & V\,e^{-i\frac{\delta m^2}{2E}L}\,V^\dagger \; +  T\,e^{-i\frac{\Delta m^2_{41}}{2E}L}\, T^\dagger\\
    & & & &\\
    (1 \times 3) & & & & (1 \times 1)
  \end{array}\right]\,, \label{Eq:A1.5} &
\end{align} 
where
\begin{align}
  \left(\begin{array}{cccc}
    N & & S \\
    & &   \\
    V & & T
  \end{array} \right)  = \left(
  \begin{array}{ccc|c}
    \widetilde{U}_{e1}    & \widetilde{U}_{e2}    & \widetilde{U}_{e3}    & \widetilde{U}_{e4}    \\
    \widetilde{U}_{\mu1}  & \widetilde{U}_{\mu2}  & \widetilde{U}_{\mu3}  & \widetilde{U}_{\mu4}  \\
    \widetilde{U}_{\tau1} & \widetilde{U}_{\tau2} & \widetilde{U}_{\tau3} & \widetilde{U}_{\tau4} \\ \hline
    \widetilde{U}_{s1}    & \widetilde{U}_{s2}    & \widetilde{U}_{s3}    & \widetilde{U}_{s4}
  \end{array}
  \right)\; {\rm and\;}
  H^m_{4 \times 4} & = \frac{1}{2E} \cdot \left[\begin{array}{cc}
    \delta m^2 & 0 \\
    0 & \Delta m^2_{41}
  \end{array}
  \right]\,.  \label{Eq:A1.6}
\end{align}

With the current knowledge of particle physics and its detection technology, we can not measure the sterile neutrino contents directly. However, using the principle of conservation in neutrino oscillation probabilities, one can infer the impact of such a sterile neutrino. Thus, it is essential to know the information about $N$, $S$, $\delta m^2$ and $\Delta m^2_{41}$ in the $3 \times 3$ block of eq.~(\ref{Eq:A1.5}), to calculate the oscillation probabilities. Now, when $t = 0$, the matrix of transition amplitude becomes an identity one. This gives rise to an important relation for the $3 \times 3$ blocks: $N \cdot N^\dagger\; + \; S \cdot S^\dagger\; = \; I$ and $V \cdot V^\dagger\; + \; T \cdot T^\dagger\; = \; 1$. In practice, $1 \times 1$ block does not provide any useful information because we don't have access to the initial and final states of a sterile neutrino. Now, for a generalized scenario where `n' number of neutrino species are considered, one may not have all the information regarding either $S$ or $\Delta M^2$ or even both, then only the partial information on the evolution of propagating neutrino will be available via $N\,e^{-i\frac{\delta m^2}{2E}L}\,N^\dagger$. Consequently, it gives rise to the non-conservation of three neutrino oscillation probabilities and the corresponding zero-length effect when $L$ has a null value. \\

\noindent
\textbf{Lower-triangular matrix formulation of NUNM:} For a generalized neutrino mixing scenario, global unitary mixing matrix $\widetilde{U}^{n\times n}$ contains $n(n-1)/2$ number of mixing angles and $(n-2)(n-1)/2$ number of physical phases. Using Okubo's prescription~\cite{Okubo:1962zzc}, the $\widetilde{U}^{n\times n}$ matrix can be constructed with the corresponding $n(n-1)/2$ rotational matrices as:

\begin{align}
  \widetilde{U}^{n\times n} & = R_{n-1\,n}\cdot R_{n-2\,n}\ldots R_{3\,n}\cdot R_{2\,n}\cdot R_{1\,n}\ldots R_{3\,n-1}\cdot R_{2\,n-1}\cdot R_{1\,n-1}\ldots R_{23}\cdot R_{13} \cdot R_{12}\,.
  \label{Eq:A1.7}
\end{align}

While keeping the last three rotational matrices apart from the rest of the matrix multiplications, we can form a lower-triangular matrix. For simplicity, lets consider $n=4$ which leads to six rotational matrices, and the $\widetilde{U}^{4 \times 4}$ can be expressed as:

\begin{align}
  \widetilde{U}^{4\times 4} & = R_{34}\cdot R_{24} \cdot R_{14} \cdot R_{23} \cdot R_{13} \cdot R_{12}\\
  & = R_{34}\cdot R_{24} \cdot R_{14} \cdot \mathbb{R}\,,
  \label{Eq:A.9}
\end{align}
where $\mathbb{R} = R_{23} \cdot R_{13} \cdot R_{12}$ and denoting $\sin\theta_{ij}\,\left(\cos\theta_{ij}\right) = s_{ij}\,\left(c_{ij}\right)$. For simplicity, considering physical phases to null values
\begin{align}
  R_{34} = \left(
  \begin{array}{cccc}
  1 & 0 & 0 & 0\\
  0 & 1 & 0 & 0\\
  0 & 0 & c_{34} & -s_{34}\\
  0 & 0 & s_{34} & c_{34}
  \end{array}
  \right),\;
  R_{24} = \left(
  \begin{array}{cccc}
    1 & 0 & 0 & 0\\
    0 & c_{34} & 0 & -s_{34}\\
    0 & 0 & 1 & 0\\
    0 & s_{34} & 0& c_{34}
  \end{array}  \right),\;
R_{14} = \left(
  \begin{array}{cccc}
  c_{34} & 0 & 0 & -s_{34}\\
  0 & 1 & 0 & 0\\
  0 & 0 & 1 & 0\\
  s_{34} &0 & 0& c_{34}
\end{array}  \right)\,,
\end{align}
\begin{align}
  R_{34}\cdot R_{24} \cdot R_{14} = 
  \left(\begin{array}{ccc|c}  
  c_{14}              & 0             & 0       & s_{14}             \\
  -s_{14}s_{24}       & c_{24}        & 0       & c_{14}s_{24}       \\
  -c_{24}s_{14}s_{34} & -s_{24}s_{34} & c_{34}  & c_{14}c_{24}s_{34} \\ \hline
  -c_{24}s_{14}c_{34} & -s_{24}s_{34} & -s_{34} & c_{14}c_{24}c_{34}
  \end{array}
  \right){\;\rm and\;} \mathbb{R} = \left(
  \begin{array}{cccc}
    U & & & 0 \\
        & & & \\
    0 & & & 1 
  \end{array}
  \right)\,.\label{Eq:A1.11}
\end{align}
\noindent
From eqs.~(\ref{Eq:A.9}) and (\ref{Eq:A1.11}), one can parameterize non-unitary active neutrino mixing matrix $N$  as $N\,=\, \alpha\cdot U$, where $\alpha$ is the lower-triangular $3\times 3$ submatrix of $R_{34}\cdot R_{24} \cdot R_{14} $ shown in the above eq.~(\ref{Eq:A1.11}) and $U$ is standard unitary PMNS matrix.

\section{Effective $\nu_\mu$ survival probability in the presence of NUNM}
\label{app:A2}
Let us consider an effective two-neutrino scenario in the $\mu\text{-}\tau$ sector, and assigning the vacuum mixing angle $\theta_{23}$ and mass-squared splitting as $\Delta m^2_{32}$. Now, considering the new physics scenario of NUNM in 2-3 block and the corresponding oscillation matrix elements as follows:
\begin{align}
  U & = \left(
  \begin{array}{cc}
    \cos\theta_{23} & \sin\theta_{23} \\
    -\sin\theta_{23} & \cos\theta_{23}
  \end{array}
  \right)\,,
  \label{Eq:A2.1.1}\\
  \nonumber \\
  M & = \left(
  \begin{array}{cc}
    0 & 0 \\
    0 & \Delta m^2_{32}
  \end{array}
  \right)\,,
  \label{Eq:A2.1.2}\\
  \nonumber \\
  \hat{\alpha} & = \left(
  \begin{array}{cc}
    1+\alpha_{22} & 0 \\
    \alpha_{32} & 1+\alpha_{33}
  \end{array}
  \right)\,,
  \label{Eq:A2.1.3}\\
  \nonumber \\
  N & = \hat{\alpha}\cdot U \nonumber \\
    & = \left(
  \begin{array}{cc}
    1+\alpha_{22} & 0 \\
    \alpha_{32} & 1+\alpha_{33}
  \end{array}
  \right) \cdot 
  \left(
  \begin{array}{cc}
    \cos\theta_{23} & \sin\theta_{23} \\
    -\sin\theta_{23} & \cos\theta_{23}
  \end{array}
  \right)\,.
  \label{Eq:A2.1.4}
\end{align}
Now for the simplicity, let us consider $\alpha_{22} = \alpha_{33} = 0$.
\begin{align}
  &N = \left[
  \begin{array}{cc}
    \cos\theta_{23} & \sin\theta_{23}\\
    \alpha_{32}\cos\theta_{23} - \sin\theta_{23} & \cos\theta_{23} + \alpha_{32}\sin\theta_{23}
  \end{array}
  \right],\label{Eq:A2.2.1}
\end{align}  
\begin{align}
  &N^\dag \cdot V_{NC} \cdot N = \nonumber \\
  &V_{NC}\; \cdot \left[
  \begin{array}{cc}
    c^2_{23} + \left(\alpha_{32}c_{23}-s_{23}\right)^2 &
    c_{23}s_{23} + \left(\alpha_{32}c_{23}-s_{23}\right)\left(c_{23}+\alpha_{32}s_{23}\right)\\
    c_{23}s_{23} + \left(\alpha_{32}c_{23}-s_{23}\right)\left(c_{23}+\alpha_{32}s_{23}\right) & s^2_{23} + \left(c_{23}+\alpha_{32}s_{23}\right)^2
  \end{array}
  \right]\,.
  \label{Eq:A2.2.2}
\end{align}
The effective Hamiltonian ($\mathcal{H}$) can be expressed as follows:
\begin{align}
  \mathcal{H} & = \frac{M}{2E} + N^\dag \cdot V_{NC} \cdot N\,,\\
  \nonumber\\
  \mathcal{H} & = V_{NC}\; \cdot \left[
  \begin{array}{cc}
    c^2_{23} + \left(\alpha_{32}c_{23}-s_{23}\right)^2 &
    c_{23}s_{23} + \left(\alpha_{32}c_{23}-s_{23}\right)\left(c_{23}+\alpha_{32}s_{23}\right)\\
    c_{23}s_{23} + \left(\alpha_{32}c_{23}-s_{23}\right)\left(c_{23}+\alpha_{32}s_{23}\right) & \Delta m^2_{32}/2E\,V_{NC} + s^2_{23} + \left(c_{23}+\alpha_{32}s_{23}\right)^2
  \end{array}
  \right]\,.
\end{align}
The eigenvalues of this effective Hamiltonian, half of the differences of these eigenvalues, will reflect the impact of new physics in the modified oscillation parameters.
\begin{align}
  \lambda_1 = \frac{1}{2}&\left.\bigg[\Delta m^2_{32}/2E + 2V_{NC} + V_{NC}\alpha^2_{32} - \bigg\lbrace\big(\Delta m^2_{32}/2E\big)^2 + 4 V^2_{NC} \alpha_{32}^2 + V^2_{NC} \alpha_{32}^4\right. \nonumber\\
  &\left.- 2\cdot\Delta m^2_{32}/2E \cdot V_{NC} \alpha_{32}^2 \cos 2\theta_{23} + 4\cdot \Delta m^2_{32}/2E \cdot V_{NC} \alpha_{32} \sin 2\theta_{23}\bigg\rbrace^{1/2}  \,\right]\,,\\
  \nonumber\\
  \lambda_2 = \frac{1}{2}&\left.\bigg[\Delta m^2_{32}/2E + 2V_{NC} + V_{NC}\alpha^2_{32} + \bigg\lbrace\big(\Delta m^2_{32}/2E\big)^2 + 4 V^2_{NC} \alpha_{32}^2 + V^2_{NC} \alpha_{32}^4\right. \nonumber\\
  &\left.- 2\cdot\Delta m^2_{32}/2E \cdot V_{NC} \alpha_{32}^2 \cos 2\theta_{23} + 4\cdot \Delta m^2_{32}/2E \cdot V_{NC} \alpha_{32} \sin 2\theta_{23}\bigg\rbrace^{1/2}  \,\right]\,, \\
  \nonumber \\
  \frac{\Delta\lambda}{2} = \frac{1}{2}&\bigg\lbrace\big(\Delta m^2_{32}/2E\big)^2 + 4 V^2_{NC} \alpha_{32}^2 + V^2_{NC} \alpha_{32}^4 - 2\cdot\Delta m^2_{32}/2E \cdot V_{NC} \alpha_{32}^2 \cos 2\theta_{23} \nonumber \\
  & + 4\cdot \Delta m^2_{32}/2E \cdot V_{NC} \alpha_{32} \sin 2\theta_{23}\bigg\rbrace^{1/2}\,.
\end{align}
Here, $\Delta \lambda =\; \lambda_2 - \lambda_1$. Now considering an approximated solution to $\Delta \lambda$ by limiting the higher ordered $\alpha^2_{32}$ terms $\simeq 0$.
\begin{align}
  \frac{\Delta\lambda}{2} & = \frac{1}{2}\bigg[\big(\Delta m^2_{32}/2E\big)^2 + 4\cdot \Delta m^2_{32}/2E \cdot V_{NC} \alpha_{32} \sin 2\theta_{23}\bigg]^{1/2}\,,\\
  \nonumber \\
  \frac{\Delta\lambda}{2} & = \Delta m^2_{32}/4E\bigg[1 + 4\cdot(\Delta m^2_{32}/2E)^{-1} \cdot V_{NC} \cdot \alpha_{32} \cdot \sin 2\theta_{23}\bigg]^{1/2}\,, \\ 
  \nonumber\\
  \frac{\Delta\lambda}{2} & \simeq \Delta m^2_{32}/4E\bigg[1 + 2\cdot(\Delta m^2_{32}/2E)^{-1} \cdot V_{NC} \cdot \alpha_{32} \cdot \sin 2\theta_{23} + \ldots\bigg]\,,  
\end{align}  
ignoring higher ordered expansion terms of $4\cdot(\Delta m^2_{32}/2E)^{-1} \cdot V_{NC} \cdot \alpha_{32} \cdot \sin 2\theta_{23}\,,$
\begin{align}
  \frac{\Delta\lambda}{2} & \simeq \frac{\Delta m^2_{32}}{4E}\, +\, V_{NC} \cdot \alpha_{32} \cdot \sin 2\theta_{23}\,,
\end{align}
\begin{equation}
  \boxed{\frac{\Delta\lambda}{2}\cdot L \simeq \bigg(\frac{\Delta m^2_{32}}{4E}\, +\, V_{NC} \cdot \alpha_{32} \cdot \sin 2\theta_{23}\bigg)\cdot L\,.}
\end{equation}
\newline
Now recalling the $N\cdot N^\dag$,
\begin{align}
  &N \cdot N^\dag = \left(
  \begin{array}{cc}
    1 & \alpha_{32}\\
    \alpha_{32} & 1 + \alpha^2_{32}
  \end{array}
  \right)\,,\\
  \nonumber\\
  &N \cdot N^\dag = \left[
  \begin{array}{cc}
    (NN^\dag)_{\mu\mu} & (NN^\dag)_{\mu\tau}\\
    (NN^\dag)_{\tau\mu} & (NN^\dag)_{\tau\tau}
  \end{array}
  \right]\,.\\
  \nonumber\\
  &\text{Thus,\,}\big(N \cdot N^\dag\big)_{\mu\mu} = 1,\nonumber\\
  \nonumber\\
  &A\left(\nu_\mu \to \nu_\mu\right) = \bigg(N \cdot e^{-i\cdot \mathcal{H} \cdot L} \cdot N^\dag\bigg)_{\mu\mu}\,, \\
  \nonumber\\
  &A\left(\nu_\mu \to \nu_\mu\right)(L) = \frac{1}{2}\cdot e^{-i\cdot \lambda_1\cdot L} \cdot \bigg[\left(1 + e ^{i\cdot \Delta\lambda\cdot L}\right) - \cos2\theta_{23}\left(1 - e ^{i\cdot \Delta\lambda\cdot L}\right)\bigg]\,.\\
  \nonumber\\
  &P\left(\nu_\mu \to \nu_\mu\right)(L) = \sin^2{2\theta_{23}} \cdot \cos^2\left(\frac{\Delta\lambda}{2}\cdot L\right)\, +\, \cos^2{2\theta_{23}}\,,\label{Eq:A1}\\
  \nonumber \\
  &P\left(\nu_\mu \to \nu_\mu\right)(L)\bigg|_{\theta_{23} = \pi/4} = \cos^2\left(\frac{\Delta\lambda}{2}\cdot L\right)\,,\\
  \nonumber\\
  &P\left(\nu_\mu \to \nu_\mu\right)(L)\bigg|_{\theta_{23} = \pi/4} = \cos^2\left[\bigg(\frac{\Delta m^2_{32}}{4E}\, +\, V_{NC} \cdot \alpha_{32}\bigg)\cdot L\right]\,,\\
  \nonumber\\
  &P\left(\nu_\mu \to \nu_\mu\right)(L = 0) = \bigg|\big(N \cdot N^\dag\big)_{\mu\mu}\bigg|^2 = 1\,.
\end{align}
\noindent
Now accounting for the suppression factor with the NUNM zero-distance effect, i.e., $P_{\rm eff} = P(L)/P(0)$. Thus, the effective survival oscillation probability of $\left(\nu_\mu \to \nu_\mu\right)$, can be expressed as follows:
\begin{equation}
\boxed{P_{\rm eff}\left(\nu_\mu \to \nu_\mu\right)(L)\bigg|_{\theta_{23} = \pi/4} = \cos^2\left[\bigg(\frac{\Delta m^2_{32}}{4E}\, +\, V_{NC} \cdot \alpha_{32}\bigg)\cdot L\right]\,.}
\end{equation}

\section{Impact of new physics phase $\phi_{32}$ on the $\nu_\mu$ survival probability}
\label{app:A3}
\noindent
In the appendix~\ref{app:A2}, we consider an effective two-neutrino scenario which is insensitive to complex phases, say $\delta_{\rm CP}$ and $\phi_{32}$. Here, we consider perturbative expansion in the new physics parameter $|\alpha_{32}|$ and the tri-bi-maximal mixing parameters using the one-mass-scale dominance approximation, i.e., $\left(\Delta m^2_{31}L/4E \gg \Delta m^2_{21}L/4E\right)$ to derive the expression of $\nu_\mu$ survival probability in three-neutrino scenario. The tri-bimaximal mixing parameters are defined as: $\sin\theta_{13}=r/\sqrt{3}$, $\sin\theta_{12}=\left(1-s\right)/\sqrt{3}$, and $\sin\theta_{23}=\left(1-a\right)/\sqrt{2}$.
Here, we consider only the impact of $|\alpha_{32}|$ and $\phi_{32}$ in $\nu_\mu$ survival channel. The expression upto the second order terms in $|\alpha_{32}|$, $r$, $s$, and $a$ is written as:
\begin{align}
P(\nu_\mu\to\nu_\mu)\, \simeq\, & \cos^2\Delta_{31} - 2|\alpha_{32}|\Delta_n\sin(2\Delta_{31})\cos\phi_{32}\,+\nonumber\\
&\frac{1}{4\Delta_{31}(\Delta_{31}-\Delta_e)^2}
\times\bigg[\Delta_{31}\bigg\{2r^2\Delta^2_{31}\cos\Delta_{31}\cos(\Delta_{31}-2\Delta_e)\nonumber\\
&-r^2\Delta^2_{31}+8a^2(\Delta_{31}-\Delta_e)^2\bigg\}-\Delta_{31}\cos2\Delta_{31}\bigg\{r^2\Delta^2_{31} +8a^2(\Delta_{31}-\Delta_e)^2\nonumber\\&+16|\alpha_{32}|^2(\Delta_{31}-\Delta_e)^2\Delta^2_n\cos^2\phi_{32}\bigg\}-2(\Delta_{31}-\Delta_e)\bigg\{r^2\Delta^2_{31}\Delta_e\nonumber\\ &+4|\alpha_{32}|^2(\Delta_{31}-\Delta_e)\Delta^2_n\sin^2\phi_{32}\bigg\}\sin(2\Delta_{31})\bigg]
\label{Eq:app3.1},
\end{align}
where $\Delta_{31}=\Delta m^2_{31}L/4E$, $\Delta_e = G_FN_eL/\sqrt{2}$, and $\Delta_n = -\,G_FN_nL/2\sqrt{2}$. Here, the terms associated with $\delta_{\rm CP}$, i.e., $\alpha_{21}$ and $\alpha_{31}$ are considered to be zero~\cite{Agarwalla:2021owd}. Therefore, at the leading order terms, the oscillation probability of $\nu_\mu$ survival channel is almost insensitive to $\delta_{\rm CP}$, and the impact of the new physics parameter drop down to real value of $\alpha_{32}$.

%=============================%
\end{appendix}
%=============================%

%=============================%
\bibliographystyle{JHEP}
\bibliography{ICAL_NUNM}
%=============================%

\end{document}